\documentclass[useAMS,usenatbib]{mn2e}
\usepackage{times}

\newcommand{\sect}{Section~}
\newcommand{\fig}{Fig.~}
\newcommand{\tab}{Table~}
\newcommand{\equ}{Eq.~}
\newcommand{\ie}{i.e.\ }
\newcommand{\eg}{e.g.\ }

\newcommand{\kms}{{\rm \,km\,s^{-1}}}
\newcommand{\msun}{\rm \,M_{\odot}}
\newcommand{\lsun}{\rm \,L_{\odot}}
\newcommand{\zsun}{\rm \,Z_{\odot}}
\newcommand{\rsun}{\rm \,R_{\odot}}
\newcommand{\kelvin}{\rm \,K}

\newcommand{\lcdm}{$\mathrm{\Lambda}$CDM }
\newcommand{\mvir}{M_\mathrm{{vir}}}
\newcommand{\mtwoh}{M_\mathrm{{200}}}
\newcommand{\mdm}{M_\mathrm{{DM}}}
\newcommand{\minnerdm}{M_\mathrm{{0.6}}}
\newcommand{\rvir}{R_\mathrm{{vir}}}
\newcommand{\rtwoh}{R_\mathrm{{200}}}
\newcommand{\vvir}{V_\mathrm{{vir}}}
\newcommand{\vmax}{V_\mathrm{{max}}}
\newcommand{\vtwoh}{V_\mathrm{{200}}}
\newcommand{\tvir}{T_\mathrm{{vir}}}
\newcommand{\tdyn}{t_\mathrm{{dyn}}}

\newcommand{\zreio}{z_\mathrm{reio}}
\newcommand{\ztohot}{F_\mathrm{Zhot}}
\newcommand{\epsdisk}{\epsilon_\mathrm{disk}}
\newcommand{\epshalo}{\epsilon_\mathrm{halo}}
\newcommand{\alphasf}{\alpha_\mathrm{SF}}

\newcommand{\mstar}{M_\mathrm{{*}}}
\newcommand{\mcold}{M_\mathrm{{cold}}}
\newcommand{\mhot}{M_\mathrm{{hot}}}
\newcommand{\meject}{M_\mathrm{{eject}}}

\newcommand{\mreheat}{M_\mathrm{{reheat}}}
\newcommand{\mzstar}{M^\mathrm{Z}_\mathrm{{*}}}
\newcommand{\zstar}{Z_\mathrm{{*}}}
\newcommand{\zcold}{Z_\mathrm{{cold}}}
\newcommand{\zhot}{Z_\mathrm{{hot}}}
\newcommand{\zeject}{Z_\mathrm{{eject}}}
\newcommand{\dmcool}{\dot{M}_\mathrm{{cool}}}

\newcommand{\dmejec}{\dot{M}_\mathrm{{eject}}}
\newcommand{\dmback}{\dot{M}_\mathrm{{back}}}

\newcommand{\dmreheat}{\dot{M}_\mathrm{{reheat}}}
\newcommand{\dmzstar}{\dot{M}^\mathrm{Z}_\mathrm{{*}}}
\newcommand{\dmzcold}{\dot{M}^\mathrm{Z}_\mathrm{{cold}}}
\newcommand{\dmzhot}{\dot{M}^\mathrm{Z}_\mathrm{{hot}}}
\newcommand{\dmzejec}{\dot{M}^\mathrm{Z}_\mathrm{{eject}}}
\newcommand{\feh}{\ensuremath{\mathrm{[Fe/H]}}}

\newcommand{\toneasn}{Type \uppercase\expandafter{\romannumeral 1}a }
\newcommand{\ttwosn}{Type \uppercase\expandafter{\romannumeral 2} }
\newcommand{\hone}{H\uppercase\expandafter{\romannumeral 1} }
\newcommand{\mhone}{M_\mathrm{{H\,\uppercase\expandafter{\romannumeral 1}}}}
\newcommand{\htwo}{H\,\uppercase\expandafter{\romannumeral 2} }
\newcommand{\molehydro}{\mathrm{H_2}}
\newcommand{\snonea}{SN\,\uppercase\expandafter{\romannumeral 1}a\,}
\newcommand{\sntwo}{SN\,\uppercase\expandafter{\romannumeral 2}\,}
\newcommand{\etasn}{\eta_\mathrm{SN}}
\newcommand{\vsn}{V_\mathrm{SN}}
\newcommand{\esn}{E_\mathrm{SN}}
\newcommand{\ehot}{E_\mathrm{hot}}
\newcommand{\caltr}{Ca \uppercase\expandafter{\romannumeral 2} }
\def\spose#1{\hbox to 0pt{#1\hss}}
\def\lta{\mathrel{\spose{\lower 3pt\hbox{$\sim$}}
    \raise 2.0pt\hbox{$<$}}}
\def\gta{\mathrel{\spose{\lower 3pt\hbox{$\sim$}}
    \raise 2.0pt\hbox{$>$}}}    
\def\magsq{{\,\rm mag\, arcsec}^{-2}}

\usepackage{graphicx}
\usepackage{color}
\title[The nature of the MW satellites]
      {On the nature of the Milky Way satellites}
\author[Y.-S. Li, G. De Lucia and A. Helmi]
       {Yang-Shyang Li$^{1}$\thanks{Email: ysleigh@astro.rug.nl}, 
         Gabriella De Lucia$^{2}$
         and Amina Helmi$^{1}$\\
         $^{1}$Kapteyn Astronomical Institute, University of Groningen, 
         P.O. Box 800, 9700 AV Groningen, the Netherlands\\
         $^{2}$ INAF - Astronomical Observatory of Trieste, via G.B. Tiepolo
         11, I-34143 Trieste, Italy}
\begin{document}

%Accepted 2009 September 29. Received 2009 September 28; in original form  2009 September 7

\pagerange{\pageref{firstpage}--\pageref{lastpage}} \pubyear{2010}

\maketitle

\label{firstpage}

\begin{abstract}
We combine a series of high-resolution simulations with semi-analytic
galaxy formation models to follow the evolution of a system resembling
the Milky Way and its satellites.  The semi-analytic model is based on
that developed for the Millennium Simulation, and successfully
reproduces the properties of galaxies on large scales, as well as
those of the Milky Way.  In this model, we are able to reproduce the
luminosity function of the satellites around the Milky Way by
preventing cooling in haloes with $\vvir < 16.7\kms$ (\ie the atomic
hydrogen cooling limit) and including the impact of the reionization
of the Universe.  The physical properties of our model satellites (\eg
mean metallicities, ages, half-light radii and mass-to-light ratios)
are in good agreement with the latest observational measurements. We
do not find a strong dependence upon the particular implementation of
supernova feedback, but a scheme which is more efficient in galaxies
embedded in smaller haloes, \ie shallower potential wells, gives
better agreement with the properties of the ultra-faint satellites.
Our model predicts that the brightest satellites are associated with
the most massive subhaloes, are accreted later ($z \lta 1$), and have
extended star formation histories, with only 1 per cent of their stars
made by the end of the reionization.  On the other hand, the fainter
satellites tend to be accreted early, are dominated by stars
with age $>$ 10 Gyr, and a few of them formed most of their stars
before the reionization was complete. Objects with luminosities
comparable to those of the classical MW satellites are associated with
dark matter subhaloes with a peak circular velocity $\gta 10 \kms$, in
agreement with the latest constraints.
\end{abstract}

\section{Introduction}
The satellites of the Milky Way (MW) are powerful touchstones for galaxy
formation and evolution theories.  Their proximity facilitates detailed
observations and characterisation of their properties and hence constrains
`near-field' cosmology. In addition, their shallow potential wells make them
more sensitive to astrophysical processes such as supernova (SN)
feedback \citep{larson74,ds86} or to the presence of a photoionization
background \citep{br92}.

Deep images have allowed the construction of colour-magnitude diagrams (CMD) of
the MW satellites, from which the star formation histories have been deduced.
These studies indicate that there is a large variety in the star formation
histories of these galaxies \citep{mateo98,dolphin05}.  The two gas-rich dwarf
irregular (dIrr) Magellanic Clouds show on-going star formation while the other
dwarf spheroidal galaxies (dSphs) are gas-deficient and show generally little
evidence for recent star formation.  Modern studies have revealed that all
satellites contain an old stellar population ($>$ 10 Gyr), which likely keeps 
the imprints of the young Universe.  

In recent years, the number of known satellites around the MW has doubled,
thanks to the discovery of very low surface brightness dwarf galaxies in the
Sloan Digital Sky Survey (SDSS)
\citep{willman05b,willman05a,belokurov06,belokurov07,zucker06b,zucker06a,irwin07,walsh07,belokurov08,belokurov09}. Since
the sky coverage of SDSS DR5 is about 1/5 of the full sky and the surface
brightness limit is about $\mu \sim 30\magsq$ \citep{koposov08},
many satellites are likely yet to be discovered in the next generations of
surveys.  E.g.\ \cite{tollerud08} have used sub-samples of the Via Lactea I
\citep{dkm07} subhaloes to conclude that the total number of MW satellites
within 400~kpc should be between $\sim$300 and $\sim$600 and dominated by
satellites fainter than $M_{V}=-5$.

The new SDSS satellites have lower surface brightness ($\mu > 27\magsq$)
compared to the classical satellites, but similar physical sizes. They have
comparable luminosities to some Galactic globular clusters, but are
significantly bigger \citep{belokurov07}.  The nature of these newly discovered
satellites is still unclear.  They could be the prolongation towards fainter
luminosities of the classical MW satellites \citep{kirby08}, tidal features
\citep[\eg Hercules dSph,][]{coleman07,sand09}, or represent a completely new
class of objects.

Kinematic modelling based on line-of-sight velocity dispersions, have
demonstrated that the classical MW satellites are dominated by dark matter
\citep{mateo93,mateo98}.  Recent studies have shown that, under the assumption
of virial equilibrium, the ultra-faint satellites have mass-to-light ratios as
high as $\sim 100-1000$, implying that these objects are the most dark matter
dominated systems known \citep[\eg][]{munoz06,sg07}.  The constraint on the
total mass inferred by velocity dispersions is still uncertain because of the
mass-velocity anisotropy degeneracy and of the small number of tracers employed
in these studies.  Recent analyses suggest that the MW satellites (including
the newly discovered SDSS satellites) have a common mass scale when considering
their innermost regions within 600 or 300~pc \citep{strigari07,strigari08}.

The cold dark matter (CDM) hierarchical paradigm successfully explains the
large scale structures of the Universe \citep{spergel07}.  Semi-analytic
(hereafter SA) galaxy formation models coupled with merger trees extracted from
$N$-body simulations, represent a useful technique to diagnose the complex
physics involved in galaxy formation, with modest computational costs. In
recent years, SA models have been proved to successfully reproduce a number of
observational measurements (\eg spatial and colour-magnitude distributions) for
galaxies seen in the local Universe and at higher redshift \citep[for a recent 
review, see][]{2006RPPh...69.3101B}.  In spite of the encouraging progress on
the large scale, however, CDM still faces severe challenges on the galaxy-scale
and below.  An example is the `missing satellites problem': namely the substructures
resolved in a galaxy-size DM halo significantly outnumber the satellites
observed around the MW \citep{klypin99,moore99}. A number of studies have
suggested that astrophysical processes such as the presence of a
photoionization background might reconcile this discrepancy
\citep[\eg][]{kwg93,bkw00,benson02,somerville02}, without invoking
modifications on the nature of the DM particles to reduce the power on small
scales of the power spectrum \citep{kl00,zb03}.

Several groups have recently attempted to model the properties of the MW and
its satellites in a (semi-)cosmological setting.  For example, \citet*{kgk04}
have analysed the dynamical evolution of substructures in high-resolution
$N$-body simulations of MW-like haloes and suggested that all the luminous dwarf
spheroidals in the Local Group are descendants of the relatively massive ($\sim
10^9\, {\rm M}_{\odot}$) high-redshift haloes, which were not significantly
affected by the extragalactic ultraviolet radiation. \citet{font06} have
successfully reproduced the observed chemical abundance pattern of the MW
stellar halo by combining mass accretion histories of galaxy-size haloes with
a chemical evolution model for individual satellites. It should be noted that
in these studies, the phenomenological recipes adopted for star formation
and feedback have been tuned to reproduce some of the properties of the
satellites in the Local Group.

\cite{benson02} have used a SA model which successfully reproduces the
present-day field galaxy luminosity function to study the properties of dwarf
satellite galaxies known at the time.  Their model calculates the influence of
reionization self-consistently, based on the production of ionizing photons
from stars and quasars, and the reheating of the intergalactic medium.  This
same model reproduces quite nicely the luminosity and size distributions, gas
content and metallicity of the classical satellites of the MW and M31.  These
authors have suggested that the surviving satellites are those which formed
while the Universe was still neutral.  This study was carried out before the
boost at the faint end of the satellite luminosity function. Although they did
predict a large number of faint satellites below $M_{V}=-5$, after
extrapolating their prediction on the luminosity-size space to the faint end,
\citeauthor{benson02}'s faint satellites tend to be too small at a given
absolute magnitude compared to the ultra-faint satellites discovered in SDSS
\citep{koposov08}.  More recent studies have turned their attention to the
ultra-faint satellites: \cite{maccio09} have used three different SA galaxy
formation models to study the satellite population of MW-like galaxies.  They
have used both analytic and numerical merging histories of MW-like haloes and
shown that all three models reproduce the luminosity function of the MW down to
$M_{V}=-2$, with a hint for a bending around $M_{V}=-5$.

In this paper we combine high-resolution simulations of a MW-like halo with a
SA galaxy formation model to investigate how various astrophysical processes
affect the formation and evolution of satellites around the Milky Way. Our
study extends the analysis presented in \cite{dlh08}, which focused on the
formation of the MW galaxy and of its stellar halo. We find that by preventing
cooling in haloes with $\tvir < 10^4$~K (the atomic hydrogen cooling limit) and
including the impact of the reionization of the Universe, our model is able to
reproduce the latest measurements of the satellite luminosity function by
\cite{koposov08}.  We show that the same model reproduces the metallicity
distribution function (MDF) of the MW satellites, by including a route to
recycle metals produced in newly formed stars through the hot phase.  Our model
satellites exhibit several scaling relations similar to those followed by the
MW satellites, such as the metallicity-luminosity and the luminosity-size
relations. The properties of the model satellites resembling the newly
discovered ultra-faint SDSS satellites appear to be sensitive to the SN
feedback recipe adopted. As we will discuss in the following, our model
suggests that the surviving satellites are generally associated with haloes whose
present-day peak circular velocity, $\vmax$, $\gta 10 \kms$, total mass
exceeded a few $10^6\msun$ at $z\sim 10-20$ and which acquired their maximum
dark matter mass well above the cooling threshold, after $z\sim 6$.

This paper is organised as follows. \sect\ref{simulation_sec} presents the
simulations used in this study, and in \sect\ref{sam_sec} we summarise our
semi-analytical galaxy formation model, emphasising the new features added to
the model presented in \cite{dlh08}.  In \sect\ref{result_sec} we present our
main results and in \sect\ref{discussion_sec} we discuss the implications of
our study.  We give our conclusions in \sect\ref{conclusion_sec}.
%
%
%
%-------------------------------------------------------------------
\section{The hybrid model of galaxy formation and evolution}

\subsection{\lcdm Simulations of a MW-like halo}
\label{simulation_sec}

We have used a series of high resolution simulations of a MW-like halo
\citep{stoehr02,stoehr06}.  We note that this is the exact GAnew series used in
previous studies by \cite{lihelmi08}, \cite{dlh08} and \cite{letter}.  We
therefore only summarise the basic properties of the simulations here, and
refer the reader to those papers for more details.  The simulations were
carried out with GADGET-2\footnote{The GAnew simulations were in fact carried out using an intermediate version between GADGET and GADGET-2.} \citep{springel01b} adopting a $\Lambda$CDM
cosmological model, with $\Omega_{0}=0.3, \Omega_{\Lambda}=0.7, h=0.7$, $\sigma_{8}(z=0)=0.9$, and
Hubble constant H$_{0}$=100$h$ km s$^{-1}$~Mpc$^{-1}$. The target MW-like halo
was simulated at four increasing resolution levels (the mass resolution was
increased by a factor $9.33$ each time). In the highest resolution simulation
(GA3new), there are approximately $10^{7}$ particles with mass
$m_{p}=2.063\times 10^5\msun$ $h^{-1}$ within the virial radius.  Each
re-simulation produced 108 outputs, equally spaced logarithmically in time
between $z=37.6$ and $z=2.3$, and nearly linearly spaced from $z=2.3$ to
present.

Merger trees for all self-bound haloes were constructed as described in detail
in \citet{dlh08}. Virialised structures were identified using the standard
friends-of-friends (FOF) algorithm and linking all particles separated by less
than $0.2$ the mean inter-particle separation.  The algorithm {\small
  SUBFIND} \citep{springel01a} was then applied to each FOF group to find the
gravitationally self-bound substructures.  Following \citet{nfw97}, we define
$\rtwoh$ as the radius of a sphere enclosing a mass, $\mtwoh$, whose average
density is 200 times the critical density of the Universe at redshift $z$, \ie:
\[\mtwoh=\frac{4\pi \rtwoh^3}{3} \cdot 200 \delta_{crit}(z) = \frac{100H(z)^2 \rtwoh^3}{G}.\]
The velocity, $\vtwoh$, is defined as the circular velocity of the halo at
$\rtwoh$ ($\vtwoh=\sqrt{G\mtwoh/\rtwoh}$).  $\mtwoh$ is directly measured
from the simulations and used to calculate $\rtwoh$ and $\vtwoh$.  In our
models, we approximate the virial properties of dark matter haloes, \eg virial
radius ($\rvir)$, virial mass ($\mvir$) and virial velocity ($\vvir$) by
$\rtwoh, \mtwoh$ and $\vtwoh$ respectively, unless otherwise explicitly stated.

Following \cite{dlh08}, we scaled down the original outputs by a factor of
$1.42^3$ for the mass and $1.42$ for the positions and velocities, in order to
have a MW-like halo with $\vtwoh \sim 150\kms$
\citep{battaglia05,smith07,xue08}.  After the scaling, the smallest resolved
subhaloes (containing 20 particles) have dark matter mass $\mdm\sim 2\times
10^{6}\msun$ in the highest resolution run (GA3new).  $\mdm$ denotes the total bound mass at $z=0$ determined by {\small SUBFIND} throughout this paper.  The present-day virial
mass and the virial radius for the MW-like halo are $\mtwoh \sim 10^{12}\msun$
and $\rtwoh=209$~kpc, respectively.
\subsection{Semi-analytic modelling}
\label{sam_sec}

We use a semi-analytical galaxy formation model to study the baryonic
properties of a MW-like galaxy and its satellites.  This model has been
developed mainly at the Max--Planck--Institut f\"ur Astrophysik and we refer to
it as the `Munich model' later in the text.  The essential ideas of any
semi-analytic model can be traced back to the works by \cite{wr78} and
\cite{wf91}, and include physical processes such as the cooling of gas, star
formation and feedback due to supernova explosions. Over the years, the Munich
model has been enriched with new `ingredients', \eg the growth of
supermassive black holes \citep{kh00}, the inclusion of dark matter
substructures \citep{springel01a}, chemical enrichment \citep{delucia04b} and
AGN feedback \citep{croton06}. The model we use in this study has been
presented in \cite{dlh08} and builds upon the model whose results have been
made publicly available \citep{dlb07}. In order to reproduce the properties of
the MW satellites, we have made a few modifications to the original model. In
the following, we refer to the model used by \cite{dlh08} as the MW-model and
to the fiducial one used in this work as the satellite-model.  Here we give a brief summary
of the physical processes implemented in our models which are crucial to the
properties of the satellites.
\subsubsection{Reionization}

Following \cite{croton06}, we make use of the results by \cite{gnedin00} who
simulated cosmological reionization and quantified the effect of photoionization on the
baryon fraction of low-mass haloes. \citeauthor{gnedin00} found that reionization
reduces the baryon content in haloes whose mass are smaller than a particular
`filtering mass' scale, that varies with redshift.  The fraction of baryons, $f_{b}^{halo}$, in a halo of mass $\mvir$ at redshift $z$, is decreased
compared to the universal baryon fraction\footnote{We use the \textit{WMAP} 3-year value $f_{b}=0.17$ \citep{spergel07}.} according to the
ratio of the halo mass and the `filtering mass', $M_{F}$:

\begin{equation}
f_{b}^{halo}(z,\ \mvir)=\frac{f_{b}}{[1+0.26M_F (z)/\mvir]^3}.
\end{equation}
For $M_{F}(z)$, we use the analytical fitting function given in Appendix B of
\cite{kgk04}.

In our fiducial satellite-model, we assume that reionization starts at redshift
$z_{0}=15$ and ends at $z_{r}=11.5$, while the original MW-model adopted
$z_{0}=8$ and $z_{r}=7$ (in both models reionization lasts for about 0.12 Gyr).
For simplicity, we refer to the reionization epoch, $\zreio$, as $z_{0}$
hereafter.  Our choice of fixing $\zreio=15$ gives the right shape and the
normalisation for the satellite luminosity function.  We discuss the dependence
of the satellites luminosity function on different assumptions for the reionization epoch
in \sect\ref{lumfn_subsec}.
\subsubsection{Cooling}
In our model, the cooling of the shock-heated gas is treated as a classical
cooling flow \citep[\eg][]{wf91}, with the cooling rate depending on the
temperature and metallicity of the hot gas. As in \cite{delucia04b}, we model
these dependences using the collisional ionisation cooling curves of
\citet{sd93}. For primordial (or low-metallicity) composition, the cooling is
dominated by bremsstrahlung (free-free) emission at high temperatures $(T \gta
10^{6}\kelvin)$ and it is most efficient at $T \sim 10^5\kelvin$ and $\sim 1.5
\times 10^4 \kelvin$ (the H and He$^{+}$ peaks of the cooling function). Line
cooling from heavy elements dominates in the $10^6$--$10^7$ 
K regime for non primordial compositions. For $T<10^4 \kelvin$, \ie below the
atomic hydrogen cooling limit, the dominant coolant is molecular hydrogen
($\molehydro$). The virial temperature of a halo can be expressed as a function
of its virial velocity as:
\begin{equation}
\tvir(z)=35.9 \Bigg(\frac{\vvir(z)}{\kms}\Bigg)^2 .
\end{equation}
Therefore, a halo with $\tvir = 10^4 \kelvin$ corresponds to a virial velocity
of $\vvir = 16.7\kms$, which is equivalent to $\mvir \sim 3\times 10^7 \msun$
when $z=15$, and $\mvir \sim 2\times 10^9 \msun$ when $z=0$.

In the MW-model, haloes with $\tvir$ lower than $10^4\kelvin$, are able to cool
as much gas as a $10^4 \kelvin$ halo with the same metallicity.  In the
satellite-model, we forbid cooling in small haloes with $\tvir < 10^4\kelvin$,
for any metallicities and at all times. This is a reasonable approximation since
molecular hydrogen is very sensitive to photo-dissociation caused by UV photons
from (first) stellar objects (\citealt{har00}, see also
\citealt{kgk04,koposov09}).
\subsubsection{Star formation and supernova feedback}
\label{sam_sf_snfb_subsec}

The star formation model used in this work is described in detail in
\cite{dlh08}, while we refer to \cite{croton06} and \cite{delucia04b} for
details on the supernova feedback models. Cold gas is assumed to be
distributed in an exponential disk and provides the raw material for star
formation, which occurs at a rate: 
\begin{equation}
\psi = \alphasf M_\mathrm{sf}/\tdyn ,
\end{equation}
where $\alphasf$ is a free parameter which controls the star formation
efficiency, $\tdyn$ is the disk dynamical time and $M_{\rm sf}$ is the gas
mass above a critical density threshold. As in \cite{dlb07} and \cite{dlh08}, we
fix $\alphasf$ at 0.03, and adopt a Chabrier IMF.  The surface density
threshold takes a constant value throughout the disk \citep[see][for details]{dlh08}.  
At each time step, $\Delta t$, we calculate the amount of
newly formed stars,
\begin{equation}
\Delta \mstar = \psi\ \Delta t.
\end{equation}

Massive stars explode as SNe and inject energy in the surrounding interstellar
medium. In our model, we do not consider the delay between star formation
and SN energy (and metals) injection, \ie the lifetime of such stars is
assumed to be zero.  The energy injection by SNe per solar mass can be
expressed as $\vsn^{2}=\etasn\cdot\esn$ where $\etasn=8.0\times
10^{-3}\msun^{-1}$ is the number of SNe per unit solar mass expected from a
Chabrier IMF\footnote{This is the fraction of stars with mass larger than $\sim
  8\msun$ per unit of stellar mass formed. Note that, since we have adopted an
  instantaneous recycling approximation, it is more appropriate to compare the
  metallicities of our model galaxies with elements synthesised by Type II
  supernovae.} and $\esn=1.0\times 10^{51}$ erg is the average energy per SN.
The energy released by SNe in the same time interval is: 
\begin{equation}
\Delta \esn = \epshalo\cdot 0.5 \vsn^2 \Delta \mstar .
\end{equation}
where $\epshalo$ represents the efficiency with which 
the energy is able to reheat disk gas.  We assume that the amount of cold gas reheated by SNe is
proportional to the newly formed stars:
\begin{equation}
\Delta \mreheat = \epsdisk\ \Delta \mstar.
\label{reheat_mass_stdsn_eq}
\end{equation}
If this gas were added back to the hot phase without changing its specific
energy, the total thermal energy would change by:
\begin{equation}
\Delta \ehot = 0.5\Delta \mreheat\vvir^{2}.
\end{equation}
If $\Delta \esn > \Delta \ehot$, SN feedback is energetic enough to eject some
of the hot gas outside the halo and we assume that: 
\begin{equation}
\Delta \meject = \frac{\Delta \esn - \Delta \ehot}{0.5\vvir^2} = \Bigg(\epshalo\frac{\vsn^2}{\vvir^2}-\epsdisk\Bigg) \Delta \mstar.
\label{eject_mass_stdsn_eq}
\end{equation}
The ejected material can be re-incorporated into the hot component associated
with the central galaxy as the halo keeps growing by accreting material from
the surroundings \citep{delucia04b,croton06}.  For the two parameters which regulate
the feedback, $\epshalo$ and $\epsdisk$, we assume the values 0.35 and 3.5
respectively, following \cite{croton06} and \cite{dlb07}.  This means that for galaxies
with $\vvir<200\kms$, $\Delta\meject>0$ for $\Delta\mstar>0$.
\subsubsection*{An alternative supernova feedback}
\label{sec:alternative}

Since dwarf galaxies have shallow potential wells, their properties are
expected to be particularly sensitive to the adopted SN feedback model. To
explore this dependency, we have also used an alternative SN feedback recipe in
addition to the `standard' recipe mentioned above. We note that this
alternative feedback recipe is equivalent to the \textit{ejection} model
described in \cite{delucia04b}, and we refer to this paper for more details on
this particular feedback model. In the \textit{ejection} model, the gas
reheated by SNe is computed on the basis of energy conservation arguments and
depends on the galaxy mass as:
\begin{equation}
\Delta \mreheat= \frac{4}{3}\epsilon\ \frac{\vsn^2}{\vvir^2}\ \Delta \mstar .
\end{equation}
Our choices of the feedback parameters ($\epsilon=0.05$ and $\vsn\sim 634\kms$)
imply that galaxies with $\vvir < 87\kms$ would have in this model more heated
mass per unit of newly formed stellar mass, compared to the standard recipe
(\equ\ref{reheat_mass_stdsn_eq}).  In this model the material reheated by
supernova explosions in central galaxies is assumed to leave the halo and to
be deposited in an `ejected' component that can be re-incorporated into the hot
gas at later times. The material reheated by SNe explosions in satellite galaxies is assumed to be incorporated directly in the hot component associated with the corresponding central galaxy. 

We will discuss later how our results for the baryonic properties of the
satellites depend on the adopted feedback scheme. 

\subsubsection{Metal recycling through the hot phase}
\label{sam_sf_metalroute}
At each time step, the masses exchanged among the four phases: $\mhot$,
$\mcold$, $\mstar$, $\meject$, (\ie hot gas, cold gas, stars and ejecta) are
updated as described in \sect 4.7 and \fig 1 in \cite{delucia04b}.  The
metallicity in each phase is denoted as $Z_{x}$ and is defined as the ratio
between the mass in metals in each component ($M_{x}^{Z}$) and the
corresponding mass ($M_{x}$) where the suffix $x$ is hot, cold, star or eject.
In the satellite-model, we include a route to recycle metals produced by newly
formed stars through the hot phase of a galaxy. The equations given in \sect
4.7 of \cite{delucia04b} modify as follows: 
\[\dmzstar =+(1-R)\cdot \psi\cdot \zcold\]
\begin{eqnarray*}
\dmzhot &=&-\dmcool\cdot\zhot + \dmback\cdot\zeject \\
        &&+\sum_{sat}(\dmreheat\cdot\zcold) + \ztohot\cdot Y\cdot\psi \\
\end{eqnarray*}
\begin{eqnarray*}
\dmzcold &=&+\dmcool\cdot\zhot - (1-R)\cdot\psi\cdot\zcold \\
         && + (1-\ztohot)\cdot Y\cdot\psi -\dmreheat\cdot\zcold \\
\end{eqnarray*}
\[\dmzejec=+\dmejec\cdot\zhot-\dmback\cdot\zeject .\]
A constant yield $Y$ of heavy elements is assumed to be produced per solar mass
of gas converted into stars. The gas fraction returned by evolved stars is
$R=0.43$, appropriate for a Chabrier IMF. In the above equations, $\dmcool$
represents the cooling rate; $\dmback$ provides the re-incorporation rate;
$\dmreheat$ is the reheating rate by SNe, and $\dmejec$ is the rate of mass
ejected outside the halo. 

For the alternative SN feedback recipe, the reheated gas is assumed to be
ejected from the cold phase directly for central galaxies. In this case, the
metallicity in the ejecta is updated as
\[\dmzejec=+\dmreheat\cdot\zcold-\dmback\cdot\zeject .\]

In the MW-model, all newly produced metals returned to the cold phase
immediately, \ie $\ztohot = 0$.  Hydrodynamical simulations by
\cite{maclowf99} suggest that metals can be blown out from small galaxies with
gas mass below $10^7\msun$ (corresponding to a halo of $\mvir=3.5\times
10^8\msun$).  In our satellite model, we assume a simple two-state value for
$\ztohot$ to account for the above mass dependence:
\[\ztohot = \left\{ \begin{array}{ll}
                      0.0 & \mbox{if $\mvir \ge 5\times 10^{10}\msun$}\\
		      0.95 & \mbox{otherwise.}
                    \end{array}
            \right. \]
This means that for galaxies with a dark matter halo with virial mass less than
$5\times10^{10}\msun$, 95 per cent of newly produced metals are deposited
directly into the hot phase in this model. 
\subsection{Treatment of satellite galaxies}

We follow the convention established along the development of the Munich model
to classify galaxies according to their association with a distinct dark matter
substructure.  The galaxy associated with the most massive subhalo in a FOF
group is referred to as `Type 0' or
central galaxy.  Other galaxies in a FOF group are usually referred to as
satellites and are further differentiated into `Type 1' galaxies, if their dark
matter subhalo is still identified, and `Type 2', when their subhalo has fallen
below the resolution limit of the simulation.  

When a galaxy becomes a satellite, the dark matter mass of the parent subhalo
is approximated using the number of bound dark matter particles given by
{\small SUBFIND}.  The disk size is fixed at the value it had just before
accretion.  In our model, only Type 0 central galaxies are allowed to accrete
the material that cools from the hot gas associated with the parent FOF group.
Satellite galaxies do not have hot and ejected components and the cooling is
forbidden, \ie $\mhot=\meject=\dmcool=0$.  When a galaxy becomes a satellite,
its hot and ejected components are transferred to the corresponding components
of the central galaxy.  As a consequence, once matter leaves the cold phase of a
satellite, it does not rejoin the (Type 1 or Type 2) satellite at a later time,
but it can be accreted onto the corresponding central galaxy.
\subsection{Model improvements and the central MW-like galaxy}
\label{subsec:sam_summary}
The values of the parameters that enter in the SA model were chosen so as to
reproduce several observations of galaxies in the local Universe, in particular
the local galaxy luminosity function and the mass and luminosity of MW-like
galaxies \citep[see][]{delucia04b,croton06,dlb07}. It is therefore not entirely
surprising that this same parameter set provides results that are in nice
agreement with observational properties of our MW galaxy \citep{dlh08}. In
order to obtain a good agreement with the observed properties of the MW
satellites, however, we found that we had to implement some slight
modifications.

In our satellite-model, we adopt the same parameter set used in \cite{dlh08}
except for the reionization epoch, $\zreio$, and fraction of metal ejected into
the hot component, $\ztohot$. \tab\ref{model_table} introduces the SA models
that we employ in this study.  In the MW-model by \cite{dlh08}, reionization
occurs at $\zreio=8$, haloes with $\tvir<10^4$~K are allowed to cool at the
same rate of a $10^4$~K halo with the same hot gas metallicity, a `standard' SN
feedback recipe is adopted and all new metals are kept in the cold gas
phase ($\ztohot=0$). Our fiducial model for the satellite galaxies corresponds
to the `satellite-model' (the fourth row in \tab\ref{model_table}). In this model,
reionization starts at $z=15$, and 95 per cent of the new metals are deposited 
directly into the hot component in galaxies with $\mvir<5\times
10^{10}\msun$.  In all models with the prefix `satellite', we forbid cooling in
haloes with $\tvir<10^4$~K.

\begin{table*}
\footnotesize
\begin{center}
\caption{Nomenclature and features of the semi-analytic models used in this
  study.} 
\label{model_table} 
  \begin{tabular}{lrccr}
  \hline
Model Name        & $z_{0}, z_r$ & cooling for $\vvir<16.7\kms$ & SN feedback &
$\ztohot$ \\ 
(1)& (2) & (3) & (4) & (5) \\
\hline
MW-model          & (8, 7) & Yes & Standard & 0.0\\
\hline
satellite-model A & (8, 7) & No  & Standard & 0.0\\

satellite-model B & (15, 11.5) & No & Standard & 0.0\\
\hline
satellite-model   & (15, 11.5) & No & Standard & (0.95, 0.0)\\
\hline
satellite-model \textit{ejection} & (15, 11.5) & No & ejection & (0.95, 0.0)\\
\hline
satellite-model \textit{combined} & (15, 11.5) & No & combined & (0.95, 0.0)\\ 
\hline
  \end{tabular}
\end{center} 

{\small Different columns list: (1) the model name; (2) the reionization
  epoch; (3) the adopted cooling recipe in haloes with $\tvir<10^4$~K; (4) the
  adopted feedback recipe; (5) the fraction of metals injected directly into
  the hot component.}
\end{table*}

Below, we briefly discuss the dependence of model results for the central
MW-like galaxy on our changes for $\zreio$, the cooling in small haloes and
metal recycling through hot phase.  The results for the satellite galaxies are
presented in \sect\ref{result_sec}.  \tab\ref{mw_para_dependence} summarises
the properties of the MW-like galaxies in different SA models at $z=0$.

\begin{table*} 
\begin{center}
\footnotesize 
\caption{Properties of the Milky Way-like galaxy in the semi-analytic models
  employed in this study, for the GA3new simulation.}
\label{mw_para_dependence} 
  \begin{tabular}{lccccccc}
  \hline
Model Name        & $\mstar$  & $M_\mathrm{{bulge}}$  & $M_\mathrm{{coldgas}}$ & $M_\mathrm{BH}$ & Log$\frac{Z_{*}}{\zsun}$& Log$\frac{Z_{b}}{\zsun}$& $M_{B}$ \\
& [$10^{10} \msun$] & [$10^{10} \msun$] & [$10^{10} \msun$] & [$10^6 \msun$] & [dex] & [dex] & [mag] \\
(1)& (2) & (3) & (4) & (5) & (6) & (7) & (8) \\ 
\hline
MW-model          & $5.73$ & $0.64$ & $1.06$ & $\,\,\,8.2$ & $-0.05$ & $-0.28$ & $-19.53$ \\
\hline
satellite-model A & $5.75$ & $0.64$ & $1.07$ & $\,\,\,8.0$ & $-0.05$ & $-0.28$ & $-19.50$ \\

satellite-model B & $5.85$ & $0.62$ & $1.11$ & $\,\,\,7.1$ & $-0.06$ & $-0.30$ & $-19.52$ \\
\hline
satellite-model   & $5.88$ & $0.63$ & $1.11$ & $\,\,\,6.9$ & $-0.06$ & $-0.35$ & $-19.52$ \\
\hline
satellite-model \textit{ejection} & $8.26$ & $2.58$ & $1.09$ & $12.8$ & $\ \ \ 0.12$ & $\ \ \ 0.00$ & $-19.08$ \\
\hline
satellite-model \textit{combined} & $5.02$ & $0.84$ & $0.95$ & $12.0$ &$-0.08$ & $-0.25$ & $-19.53$ \\
\hline
  \end{tabular}
\end{center}
{\small Different columns list: (1) the model name; (2) stellar mass; (3) mass
  of the bulge; (4) cold gas content; (5) mass of the black hole; (6) logarithmic value of the total
  stellar metallicity; (7) logarithmic value of the bulge metallicity; (8)
  $B$-band absolute magnitude corrected for internal dust attenuation.}
\end{table*}

The results corresponding to the MW-model are given in the first row \citep[see also \fig 2 of][]{dlh08}.  The only difference between the MW-model and the
satellite-model A is the suppression of cooling in small haloes.  Comparing the
results of these two models, we find no significant changes in the properties
of the present-day MW-like galaxy.  In satellite-model B, we also change
$\zreio$ to 15 and keep $\ztohot=0$.  The only significant effect of an early
reionization is to bring down the black hole mass by $\sim 15$ per cent
compared to the value obtained in the MW-model.  The early reionization also results
in a slight increase of the stellar mass but this is still well within the observational
uncertainties ($\mstar \sim 5-8 \times 10^{10}\msun$).

Our fiducial satellite-model gives a total stellar mass similar to
that obtained from the MW-model and in agreement with current
observational constraints.  The results of these two models are also
very close in terms of the mass of the bulge and the cold gas content.
The black hole mass $M_\mathrm{BH}= 6.9\times 10^6\msun$ from the
satellite-model is in marginal agreement with the latest measurement
of the MW black hole mass $M_\mathrm{BH}=(4.5\pm 0.4) \times
10^6\msun$ \citep{ghez08}.  The ejection of metals into the hot
component in small galaxies (cf.\ satellite-model B) only makes the bulge
slightly more metal-poor ($\sim 0.06$ dex).  Results listed in
\tab\ref{mw_para_dependence} show that the modifications discussed
above influence only the properties of dwarf galaxies, while
preserving the properties of the MW-like galaxies discussed in
\cite{dlh08}.

The 5th row in Tables \ref{model_table} and \ref{mw_para_dependence}
corresponds to a model which incorporates the `alternative' (or
\textit{ejection}) SN feedback scheme described in
\sect\ref{sam_sf_snfb_subsec}.  When compared to the `standard' scheme, the dependency of the amount of reheated gas on
$1/\vvir^{2}$ in this scheme results in a more efficient feedback for
small galaxies and in a less efficient ejection for more massive
systems like the MW galaxy. These galaxies tend to have larger stellar
masses, a more massive bulge and tend to be more
metal-rich in this scheme.  As we will show in \sect\ref{result_sec}, however, this
alternative feedback scheme is able to better match the properties of
the MW satellites.

We therefore propose a combination of these two feedback recipes to account for the properties of galaxies on large and small mass scales.  This model corresponds to the `satellite-model {\it combined}', and in
this scheme, we calculate the amount of gas reheated by SNe depending on the
local potential well (mass of the associated subhalo):

\[\mreheat = \left\{ \begin{array}{ll}
  \frac{4}{3}\epsilon\ \frac{V_\mathrm{SN}^2}{\vvir^2}\ \Delta \mstar &
  \mbox{if $\vvir^2 <
    \frac{4}{3}\frac{\epsilon}{\epsdisk}\ V_\mathrm{SN}^2$}\\ 
  \epsdisk\Delta \mstar & \mbox{otherwise.}
                    \end{array}
            \right. \] 
The reheated gas is treated as in the \textit{ejection} scenario, \ie it is
added to the ejected component of a central galaxy and lost into the hot
component for a satellite.  The `satellite-model \textit{combined}' entry in
\tab\ref{mw_para_dependence} lists the properties of the MW-like galaxy in this
scheme. As expected, the stellar mass, the total and bulge metallicity, as well
as the total luminosity are now very similar to what we get with the standard
feedback scheme, albeit the bulge (and the black hole) are now more
massive.  The last two models listed in \tab\ref{model_table} populate the
same set of subhaloes with stars, with almost identical properties.  For
simplicity, we will not discuss the results of the \textit{combined} scheme
for satellite galaxies. 
%
%
%----------------------------------------------------------------------------
\section{The Milky Way satellites}
\label{result_sec}

In this Section, we define as SA model satellites of the MW-like galaxy those
that satisfy the following conditions at $z=0$: (i) a satellite has to belong 
to the same FOF group where the MW-like galaxy is; (ii) the distance to the
MW-like galaxy must be $< 280$~kpc; (iii) the galaxy is associated with a dark
matter subhalo (\ie it has to be Type 1 galaxies).  The distance cut
corresponds to the current observational limits, but we include in our
comparison the very distant and recently discovered satellite Leo T (at $\sim
420$~kpc from the Milky Way).  

We do not consider here satellites that had their dark matter haloes tidally
stripped below the resolution limit of the simulation (Type 2 galaxies). This
selection is motivated by the fact that while resolved, our Type 1 galaxies are dark matter dominated at all radii \citep[see \sect\ref{dmmass_sect}, and in agreement with observations of satellites around the MW and M31, \eg][]{sg07,strigari07,strigari08,walker09}.  Since Type 2
galaxies in our fiducial model, reach a maximum virial mass during
their evolution larger than $ 6.8\times 10^7 \msun$, this means that these
galaxies have lost more than $97$ per cent of their dark matter by present.  Therefore it
is quite unlikely that a bound stellar core would survive such a severe tidal
stripping.  We will discuss more about Type 2 model galaxies in
\sect\ref{type2_and_uf_subsec}.  Throughout this paper, we will refer to
subhaloes that host stars as `luminous satellites' or simply `satellites', and
will refer to those that do not host any star as dark satellites or subhaloes.
\subsection{The satellite luminosity function}
\label{lumfn_subsec}

Our fiducial satellite-model gives 51 luminous satellites within
280~kpc for GA3new.  This is in good agreement with the
estimated `all sky' number of satellites ($\sim 45$) brighter than
$M_{V}=-5.0$ by \cite{koposov08}.  If we remove the distance
constraint, the number of satellites is only increased by one, and the
number of subhalos varies from 1865 to 1869.

\begin{figure}
\centerline{\includegraphics[width=0.5\textwidth]{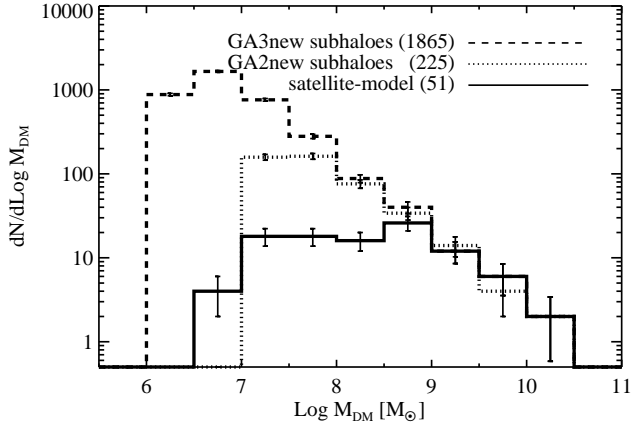}}
   \caption{The solid histogram shows the present-day mass function of
model satellites using the satellite-model for the highest resolution simulation.  Error
bars denote the 1-$\sigma$ Poisson uncertainties. The dashed histogram
shows the subhalo mass function for the highest resolution simulation,
which steeply rises up to the resolution limit.  The dotted histogram
is the subhalo mass function for a lower resolution simulation
 (\ie GA2new).} 
\label{massfn_res_fig}
\end{figure}

The mass functions for the fiducial satellite-model and of the
surviving subhaloes within 280~kpc in GA3new are shown in
\fig\ref{massfn_res_fig}.  The mass plotted here is the dark matter
mass at $z=0$ determined by {\small SUBFIND}. As indicated by the
dashed histogram, all subhaloes with present-day $\mdm > 10^9\msun$
resolved in GA3new host luminous satellites.  The mass function of
these subhaloes deviates from the power-law shape mass function of the
full subhalo population and is fairly flat below $\mdm = 10^9\msun$,
down to the resolution limit ($\mdm \sim 10^{6.5}\msun$).  For
comparison, the dotted histogram in \fig\ref{massfn_res_fig}
shows the mass function of surviving subhaloes, within the same
distance range from the central galaxy, from the lower resolution
simulation GA2new.  The smallest subhalo which could be resolved in
GA2new has $\mdm\sim 2 \times 10^7\msun$.  The subhalo mass functions
from the two simulations agree well down to $10^8\msun$.  At lower
masses, numerical effects start to become important for GA2new.
However, the number of satellites with $\mdm < 10^9\msun$ is still
lower than the number of subhaloes resolved in GA2new, which suggests
that numerical resolution should not be an issue and that the observed
decline of luminous satellites is a result of how we model the
baryonic physics, \eg SN feedback, as we will see later.

\begin{figure}
\centerline{\includegraphics[width=0.5\textwidth]{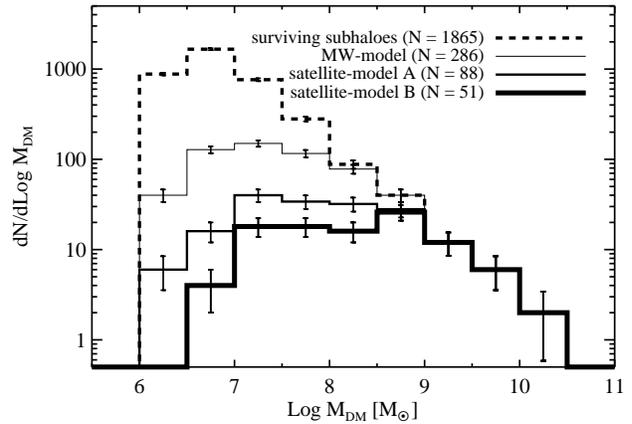}}
   \caption{Present-day mass functions (solid histograms) for
   dark matter subhaloes associated with satellites predicted by the
   semi-analytic models listed in \tab\ref{model_table}.  The dashed
   histogram again indicates the subhalo mass function for the GA3new
   simulation.}
\label{massfn_fig_tab1}
\end{figure}

\fig\ref{massfn_fig_tab1} compares the mass function for the SA
models explored in this study. In \tab\ref{sat_para_dependence}, we
list the number of luminous satellites for these models.  Note that
all the models which adopt $\zreio=15$ and forbid cooling in haloes
with $\vvir<16.7\kms$, \ie satellite-model B, satellite-model,
satellite-model \textit{ejection} and satellite-model
\textit{combined}, populate galaxies in the same set of 51 subhaloes.
The mass functions of these four models are therefore identical.

\begin{table}
\footnotesize
\begin{center}
\caption{Number of satellites around the model MW galaxy for the
  different SA models used in this study for the GA3new simulation.}
\label{sat_para_dependence} 
  \begin{tabular}{lrcr}
  \hline
Model Name        &$N_\mathrm{{sat}}$  & $z_{0}, z_r$ & $\ztohot$ \\
%& & & \\
(1)& (2) & (3) & (4) \\
\hline
MW-model          & 286 & (8, 7) & 0.0\\
\hline
satellite-model A &  88 & (8, 7) & 0.0\\

satellite-model B &  51 & (15, 11.5) & 0.0\\
\hline
satellite-model   &  51 & (15, 11.5) & (0.95, 0.0)\\
\hline
satellite-model \textit{ejection} &  51 & (15, 11.5) & (0.95, 0.0)\\
\hline
satellite-model \textit{combined} &  51 & (15, 11.5) & (0.95, 0.0)\\ 
\hline
  \end{tabular}
\end{center} 

{\small Different columns list: (1) the model name; (2) the number of luminous
  satellites; (3) the adopted reionization epoch; (4) the fraction of metals
  ejected directly into the hot component.}
\end{table}

\begin{figure}
\centerline{\includegraphics[width=0.5\textwidth]{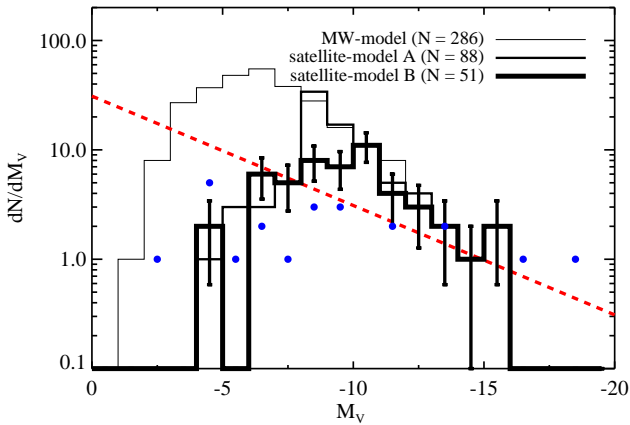}}
   \caption{Luminosity function of the observed MW satellites and for the model
     satellites around a MW-like galaxy, for the first three models listed in
     \tab\ref{sat_para_dependence}.  The integrated $V$-band luminosities of MW
     satellites (dots) are taken from various sources.  The classical dSphs are
     from \protect\cite{mateo98}; most of the ultra-faint dwarfs are from
     \protect\cite{martin08}, except Leo T \protect\citep{ryanweber08} and Leo
     V \protect\citep{belokurov08}.}
\label{lumfn_fig_tab1} 
\end{figure}

\fig\ref{lumfn_fig_tab1} shows the luminosity functions of the first
three SA models listed in the tables. The (red) dashed line shows the
`all sky SDSS' power-law luminosity function of satellites within
280~kpc, estimated by \cite{koposov08}.  For satellite-model B, we
show the 1-$\sigma$ Poisson noise. The filled circles show the
luminosity function of the 22 known satellites of the MW, including
the latest ultra-faint satellite Leo V \citep{belokurov08}. As a
reference, in the MW-model the number of surviving satellites is 286.
The drastic difference between the MW- and the satellite-model B is
mostly at the faint end of the luminosity function, and it is due to
the combined effect of an early reionization and no cooling in small
haloes.  In a model with no cooling in small haloes and a later
reionization, the number of satellites is 88 (compare the MW-model and
the satellite-model A).  The number of satellites is further reduced
to 51 when assuming an earlier reionization epoch ($\zreio=15$), with
a reduction of galaxies in the luminosity range $M_{V} \in [-7, -10]$
compared to a model with a later reionization.  We have also
experimented values for $\zreio$ at redshift 10 and 12 and found a
number of surviving satellites of $N_\mathrm{{sat}}=79$ and $58$,
respectively.  To summarise, these choices of $\zreio=[8,10,12,15]$
all give a number of satellites down to $M_{V}=-5$ which is consistent
with the estimation by \cite{tollerud08} (see their \fig6).  However,
after examining the shape of the luminosity functions with different
$\zreio$, we decided to use $\zreio=15$ in our fiducial model, as this
choice gives both the right normalisation and a shape which is in
better agreement with the observational measurements (see next
paragraphs). We recall that our results are based on only one halo,
with mass comparable to that estimated for our Milky Way. A certain
scatter in model predictions, due \eg to the assembly history of the
parent halo, is expected and this might be significantly larger than
the Poisson noise plotted in \fig\ref{lumfn_fig_tab1}.

\begin{figure*}
\bigskip
\centerline{\includegraphics[width=0.5\textwidth]{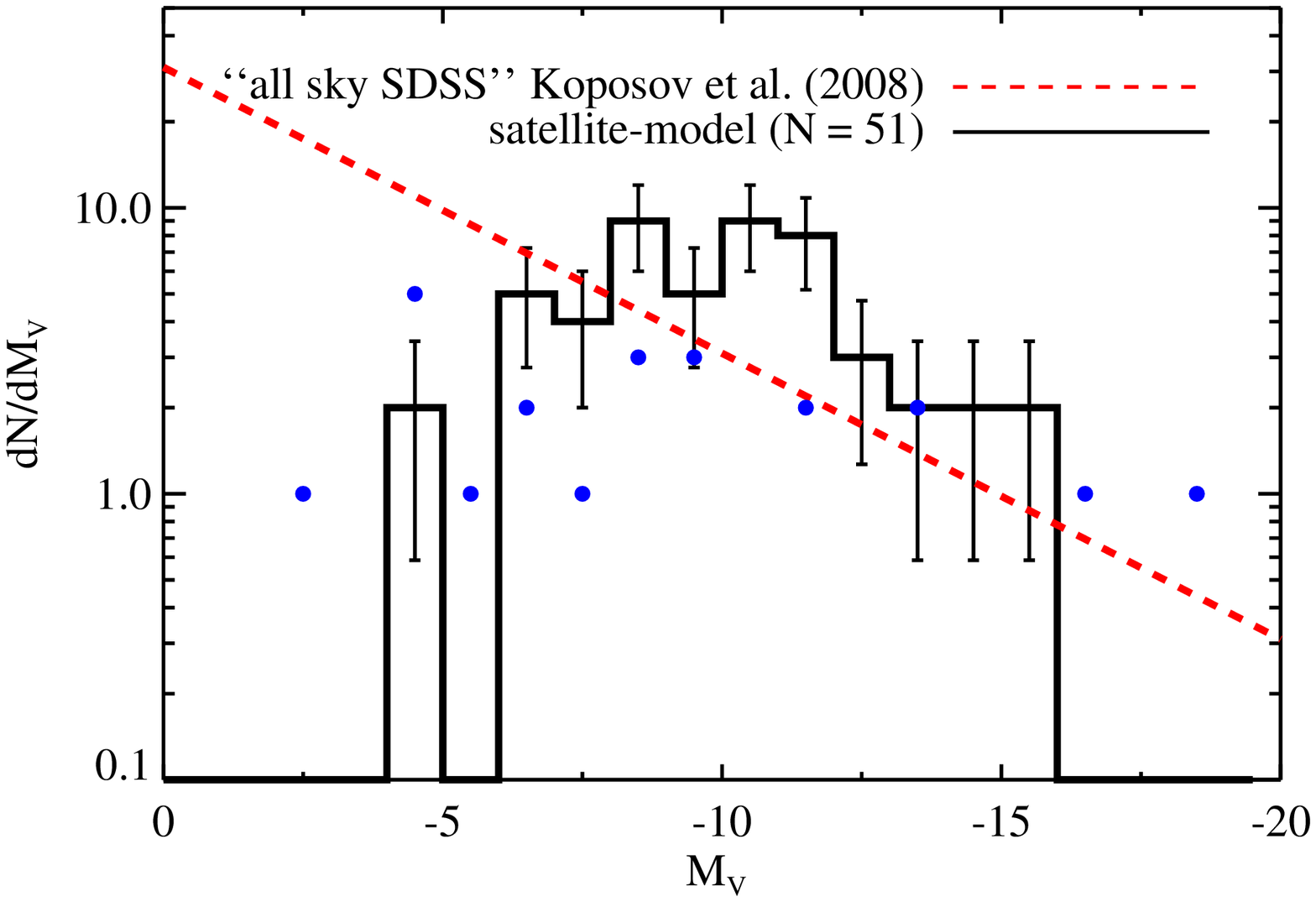}\includegraphics[width=0.5\textwidth]{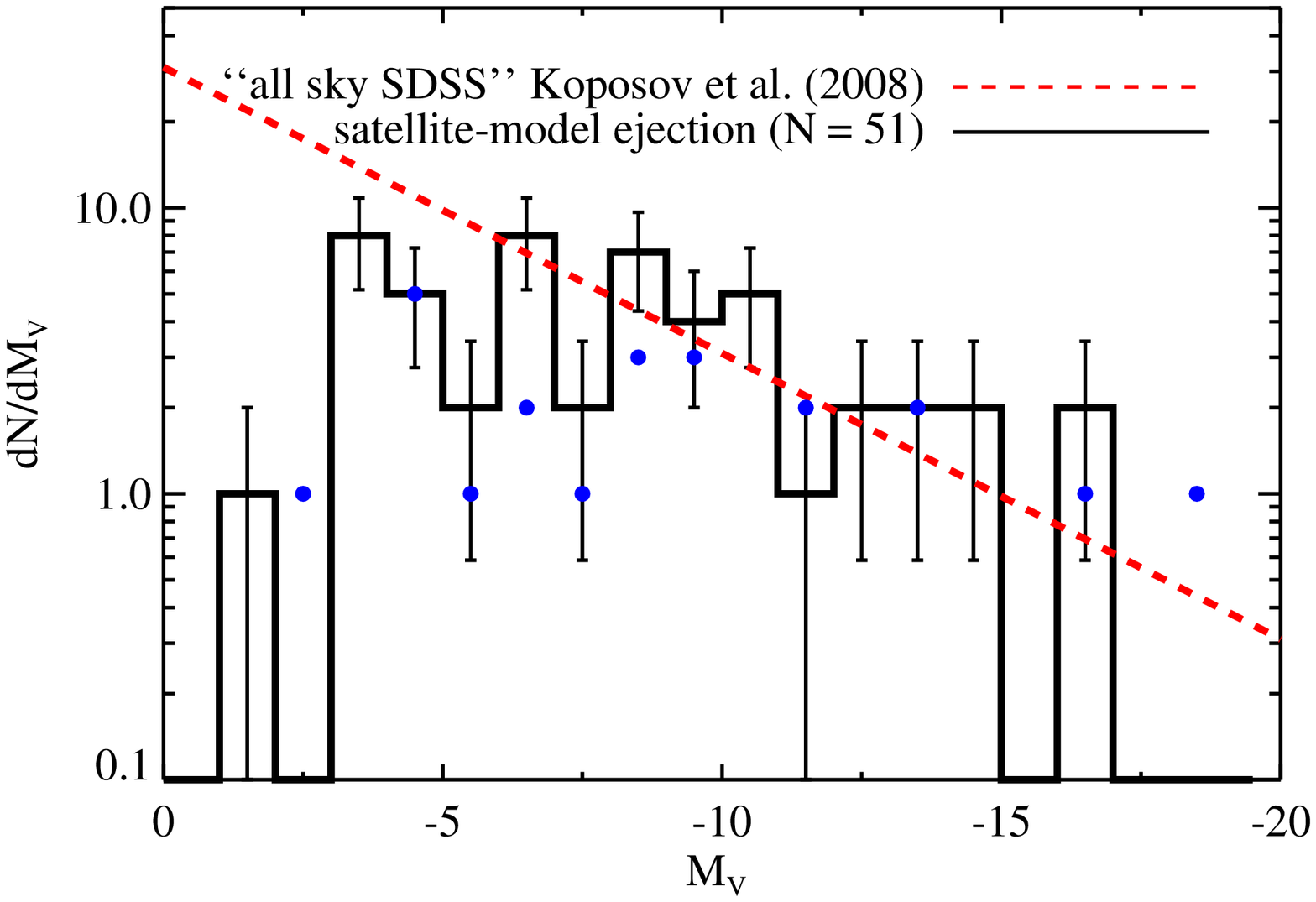}}
   \caption{Luminosity function for the observed MW satellites (filled circles
     and red dashed line) and for the model satellites (black histograms).  The
     left panel is for our fiducial model and the right panel is for the model
     with a more efficient SN feedback for small galaxies (see
     \sect\ref{sam_sf_snfb_subsec}).  Data are the same as those in
     \fig\ref{lumfn_fig_tab1}.}
\label{lumfn_fig}
\end{figure*}

The left panel in \fig\ref{lumfn_fig} shows the luminosity function of
the satellites in our fiducial satellite-model (solid histograms),
compared to the `all sky SDSS' luminosity function by \cite{koposov08}
and the observational data for the 22 known MW satellites. The model
luminosity function covers a similar luminosity range as the 22 MW
satellites though the model predicts a lower number of faint ($M_{V}>
-5$) satellites with respect to the expected `all sky' luminosity
function.  In fact, the satellite-model does not predict any satellite
fainter than $M_{V}= -4$ mag.  On the other hand, it shows an excess
with respect to the data of 10 $-$ 15 satellites around $M_{V}= -10$.
These satellites in excess are all within a distance of 200~kpc from
the host galaxy and have half-light radii between 100 and 500~pc. They
should therefore have appeared in the `all-sky' SDSS luminosity
function if they existed (provided the distribution for these bright
satellites is isotropic, as assumed by \citealt{koposov08}).  It also
seems that the model under-predicts the number of very bright
satellites. This is, however, the regime where both the data and the
simulations suffer from small number statistics.  

A comparison to \fig\ref{lumfn_fig_tab1} shows that the
luminosity function given by satellite-model B and by satellite-model
are very similar, implying that satellite luminosities do not depend
strongly on the fraction of newly produced metals put into the hot
component of a galaxy. However, the metallicity distribution functions
differ significantly (see \sect\ref{metaldis_sect}).

The right panel of \fig\ref{lumfn_fig} shows the luminosity function
of surviving satellites resulting from the alternative feedback scheme
(satellite model \textit{ejection}).  This model luminosity function
agrees very well with the observations.  When compared with that from
the standard feedback scheme, the alternative luminosity function
extends to fainter luminosities, reaching $M_{V} \sim -3$ and does not
show any excess at $M_{V} \sim -10$. Note that in the satellite-model
\textit{ejection}, the same 51 subhaloes are populated with luminous
galaxies. This suggests that the SN feedback alone is unlikely to
solve the `missing satellites problem' \citep{somerville02}, and that
the presence/absence of a luminous galaxy within a dark matter
substructure is due to the particular assembly and dynamical history
of the halo, and to the reionization history of the Universe.

These different results are entirely due to the SN scheme adopted. In
the {\it ejection} scheme, the amount of gas reheated by SNe scales as
$1/\vvir^2$, and for a galaxy with $\vvir < 87 \kms$, more gas is
heated by per unit of newly formed stars with respect to the standard
scheme, where the amount of gas heated by SNe is only proportional to
the amount of newly formed stars.  In this latter case, for galaxies
with low star formation rates (as is the case for galaxies that live
in small subhaloes), only very little gas is heated.  As long as the
density threshold for star formation is met, these galaxies keep
forming stars.  This is most likely the reason why our fiducial model
exhibits an excess of satellites around $M_{V}\sim -10$ and a lack of
ultra faint objects in the luminosity function.  In the alternative
feedback model, when star formation occurs in a central galaxy, some
cold gas is ejected outside the halo, and has to wait several
dynamical time-scales to be reincorporated into the hot halo, delaying
any subsequent star formation.  Furthermore, the impact on the star
formation of a satellite galaxy associated with a subhalo with $\vvir
< 87 \kms$ is more drastic in this model, because the ejected gas
cannot be re-incorporated onto the satellite any longer.  Small
satellites therefore consume their cold gas reservoir more efficiently
in this \textit{ejection} scenario, which causes the surface densities
of the cold gas to fall below the criterion for forming stars.

Given the reasonable match of the satellite-model and of
satellite-model \textit{ejection} to the observed luminosity
functions, in the rest of the paper we will concentrate on the
properties of the satellite populations in these two SA models.

\subsection{The metallicity distribution}
\label{metaldis_sect}

\begin{figure*}
\centerline{\includegraphics[width=0.5\textwidth]{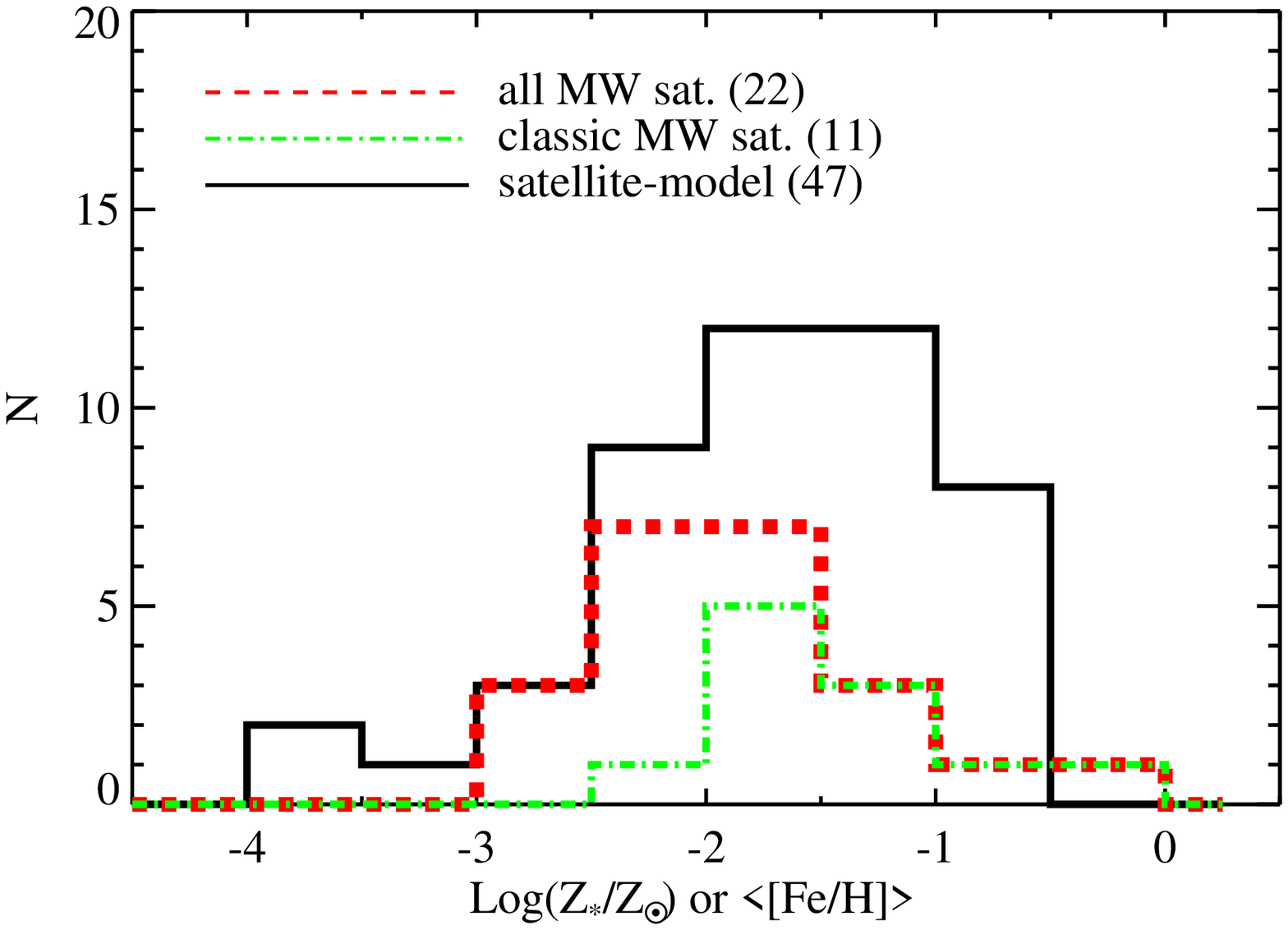}\includegraphics[width=0.5\textwidth]{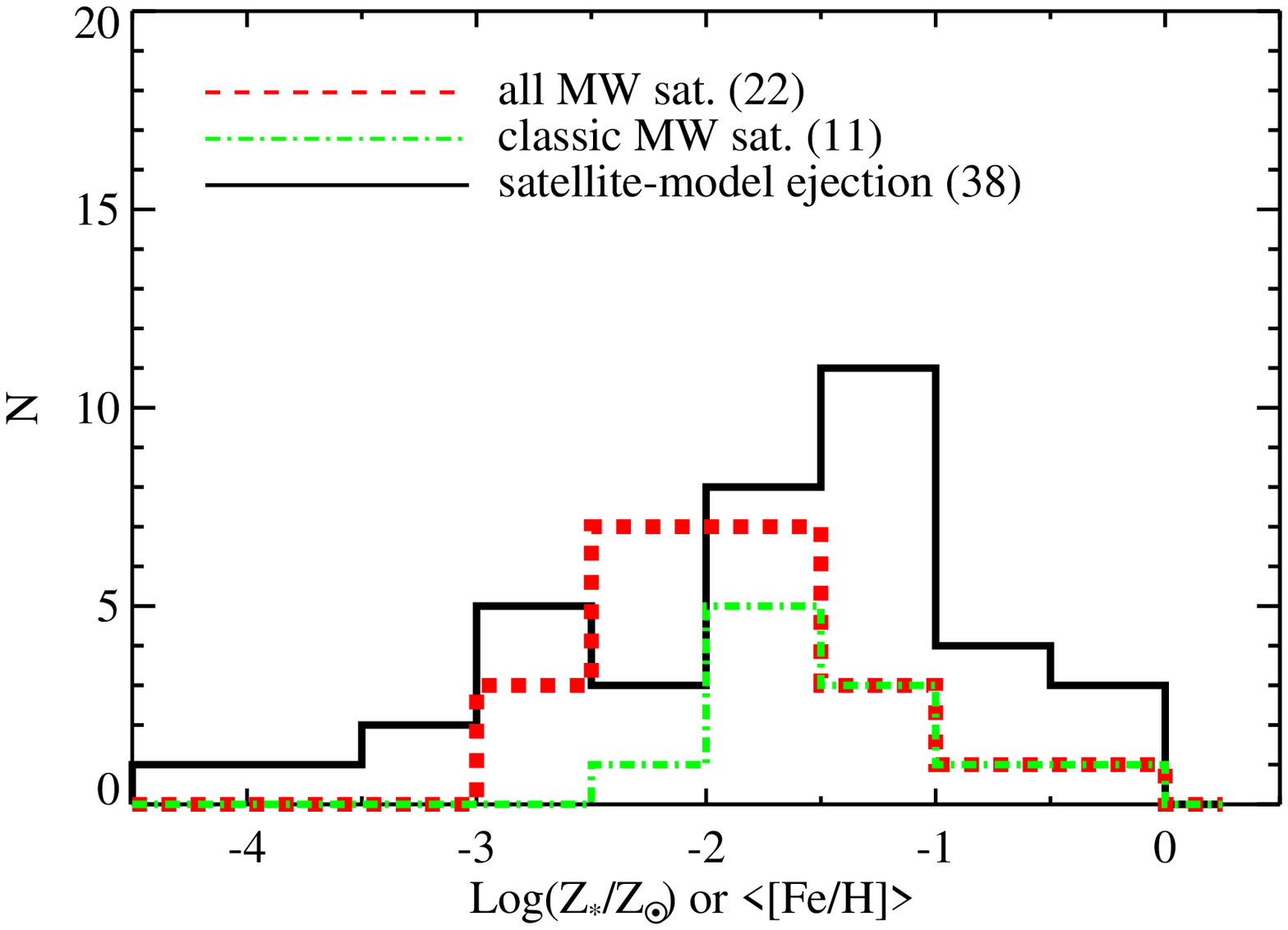}}
\caption{Histogram of the mean iron abundance $\feh$ determined for red giant
  branch stars in the MW satellites. For the model satellites we plot $\log
  (\zstar/\zsun)$.  The left panel compares the MW satellites with model
  satellites from our fiducial model and the right panel from the model with a
  more efficient SN feedback for dwarf galaxies (see
  \sect\ref{sam_sf_snfb_subsec}).  Data for the MW satellites are taken from
  various sources: LMC and SMC from \protect\cite{westerlund97}; Sgr from
  \protect\cite{cole01}; Ursa Minor and Draco from \protect\cite{harbeck01};
  Sextans, Sculptors, Carina and Fornax from the DART survey
  \protect\citep{helmi06}; Leo II from \protect\cite{koch07a}; Leo I from
  \protect\cite{koch07b}.  For the newly discovered SDSS ultra-faint dwarfs, we
  take the measurements from \protect\cite{kirby08}, except for B\"{o}otes I
  for which we use \protect\cite{munoz06}, B\"{o}otes II from
  \protect\cite{koch09} and Leo V from \protect\cite{walker_leov}.}
\label{metalfn_fig}
\end{figure*}

The left panel of \fig\ref{metalfn_fig} compares the metallicity distribution
of model satellites in our fiducial satellite-model with the observed
distribution.  The metallicity in our model is mass-weighted and is defined as
the ratio between the mass of metals in stars and total stellar mass:
\[\zstar=\mzstar/\mstar .\]
Since our model does not distinguish the long-lived main iron contributors
(\snonea) from the short-lived $\alpha$-elements enrichers (\sntwo), a direct
comparison of $Z$ to $\feh$ is questionable.  Nevertheless, here we assume that
the logarithmic value of the mass-weighted metallicity normalised to the solar
value ($\rm Z_{\odot}=0.02$) can be compared qualitatively with the $\feh$
derived from spectra of Red Giant Branch stars in the MW satellites.

Among the 51 surviving satellites in the satellite-model, four of them
are free of metals since they have only made stars once from pristine
gas.  These four metal-free satellites all have present-day $\mdm \lta
10^9\msun$ and $\mstar$ of $10^4 - 10^5 \msun$.  We do not include
these `metal-free' satellites in the left panel of
\fig\ref{metalfn_fig} and (metallicity) related discussions.
\fig\ref{metalfn_fig} shows the histograms of the mean $\feh$ of
resolved stars in each MW satellite.  The distribution of the 11
classical MW satellites is shown by the dotted-dashed histogram, while
the corresponding distribution including also 11 ultra-faint
satellites is given by the dashed histogram.  The metallicity
distributions of model and MW satellites cover similar ranges.
However, the peak of the metallicity distribution for the 22 MW
satellites is shifted to lower values with respect to the
corresponding distribution from the satellite-model.  The excess of
model satellites in the range of $-2<\log(\zstar/\zsun) <-0.5$
corresponds to the bump at $M_{V}\simeq -10$ seen in the luminosity
function.  

If we had set $\ztohot=0$ (as in satellite-model B), the
predicted metallicity distribution would be shifted towards even
higher metallicity. In this case, 36 satellites are more metal-rich
than $\log(\zstar/\zsun)=-1$, and only 11 of them have
$\log(\zstar/\zsun)<-1$, inconsistent with the observations. This is
why our fiducial model is preferred to satellite-model B.

In the right panel of \fig\ref{metalfn_fig} we plot the metallicity
function of the 38 non metal-free\footnote{\ie 13 satellites are
`metal-free' in this case.}  satellites in the satellite-model
\textit{ejection}.  The peak of this distribution is also shifted to
slightly higher values than observed, but it has a more even
distribution compared to the standard recipe. As discussed in
\sect\ref{lumfn_subsec}, this is due to the fact that satellites with
$\vvir < 87 \kms$ loose more of their cold gas reservoir (and a larger
fraction of their metals, see \sect\ref{sam_sf_metalroute}) in this
model with respect to the standard feedback scheme.
\subsection{Star formation histories}
\label{sf_sect}

\begin{figure*}
{\large $-16<M_{V}<-13$}
\centerline{\includegraphics[width=0.45\textwidth]{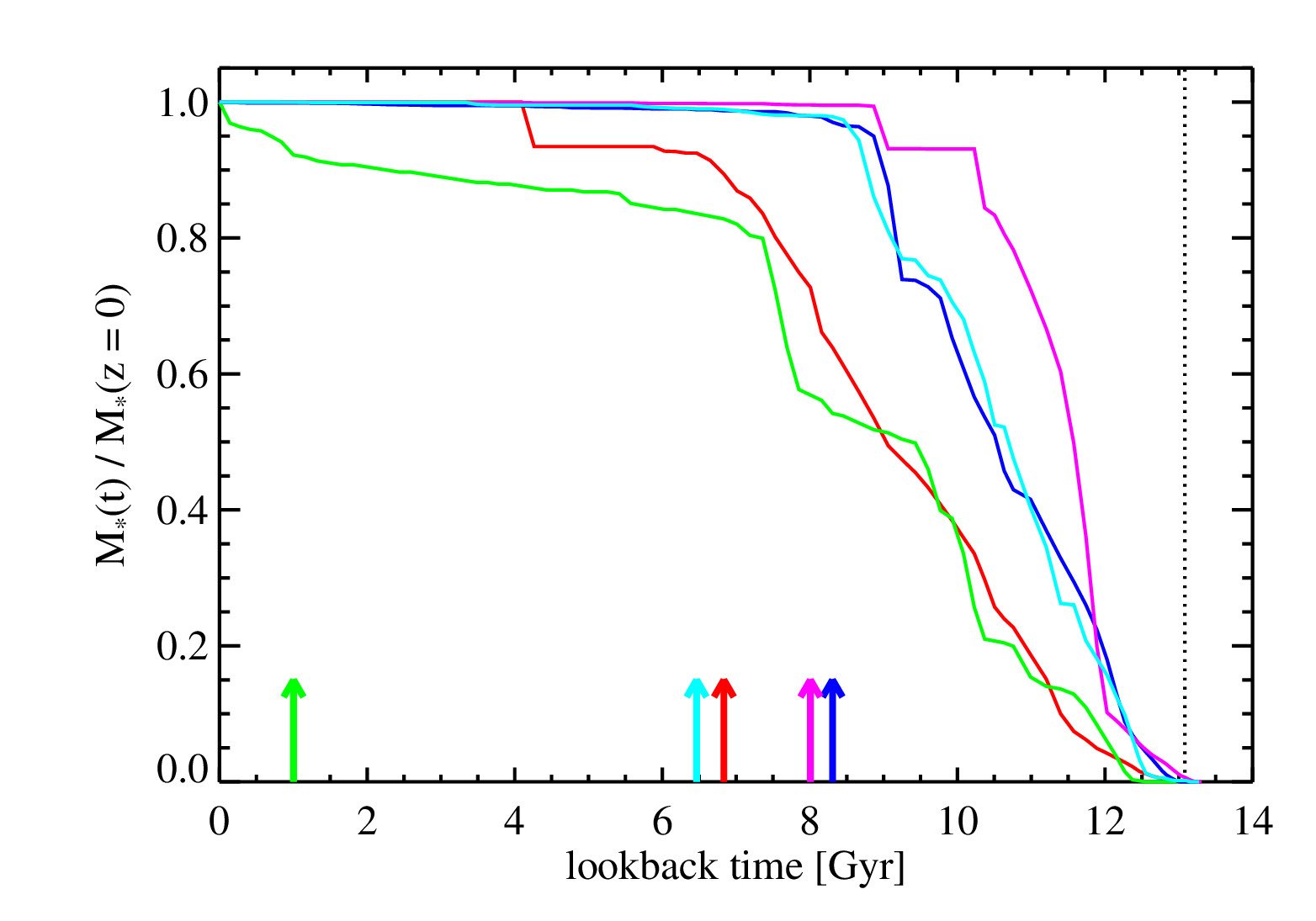}\includegraphics[width=0.45\textwidth]{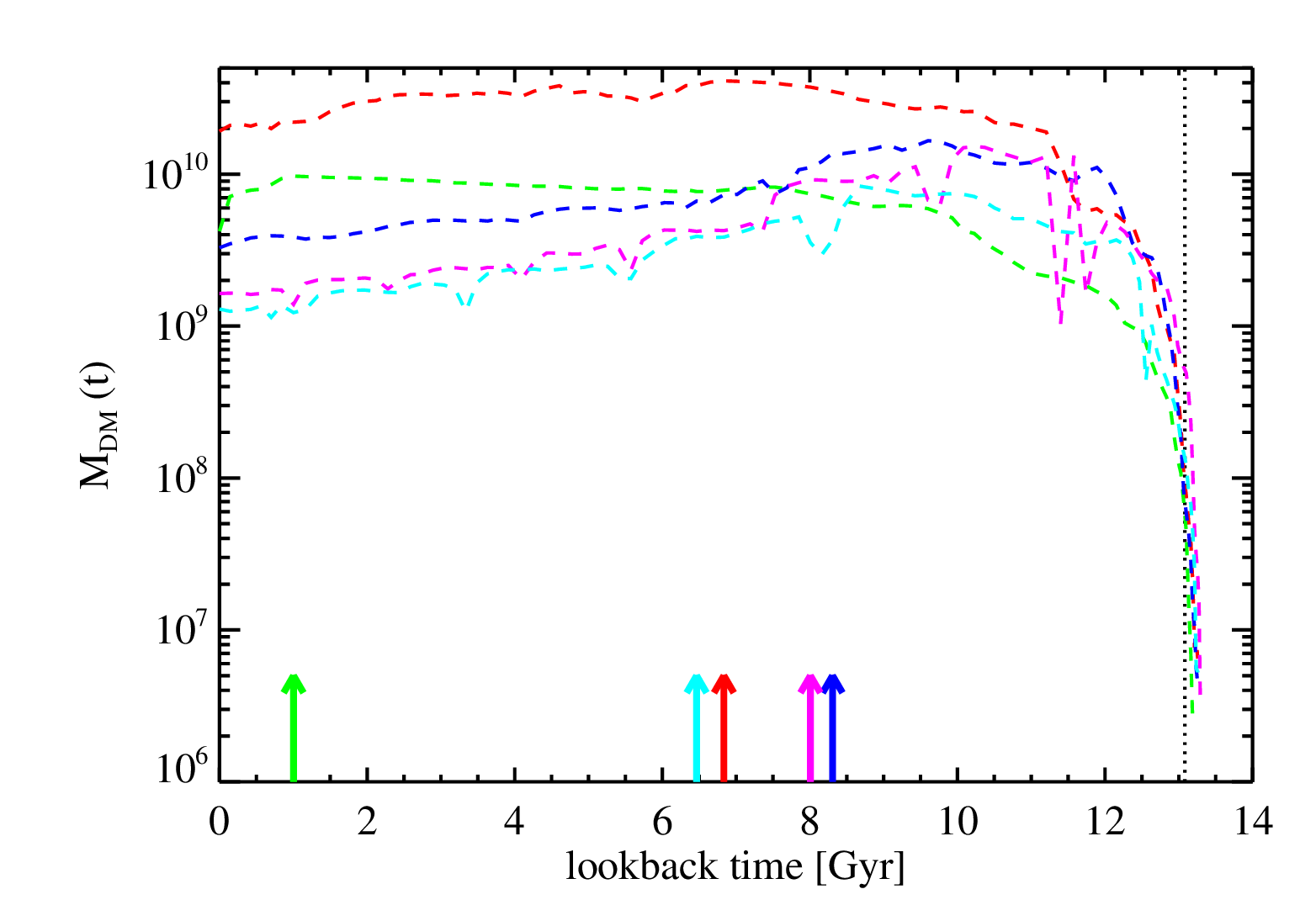}}

{\large $-12<M_{V}<-10$}
\centerline{\includegraphics[width=0.45\textwidth]{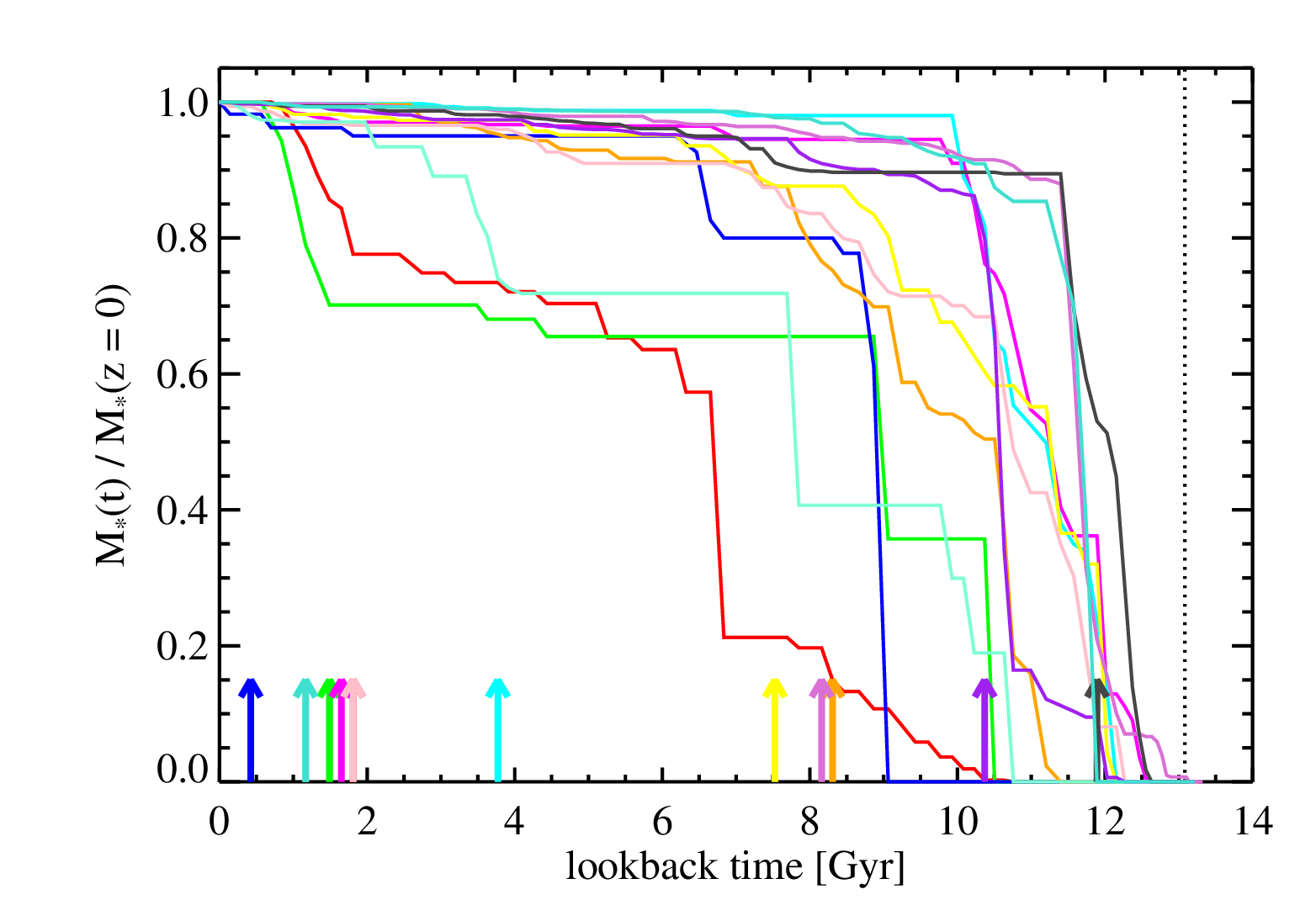}\includegraphics[width=0.45\textwidth]{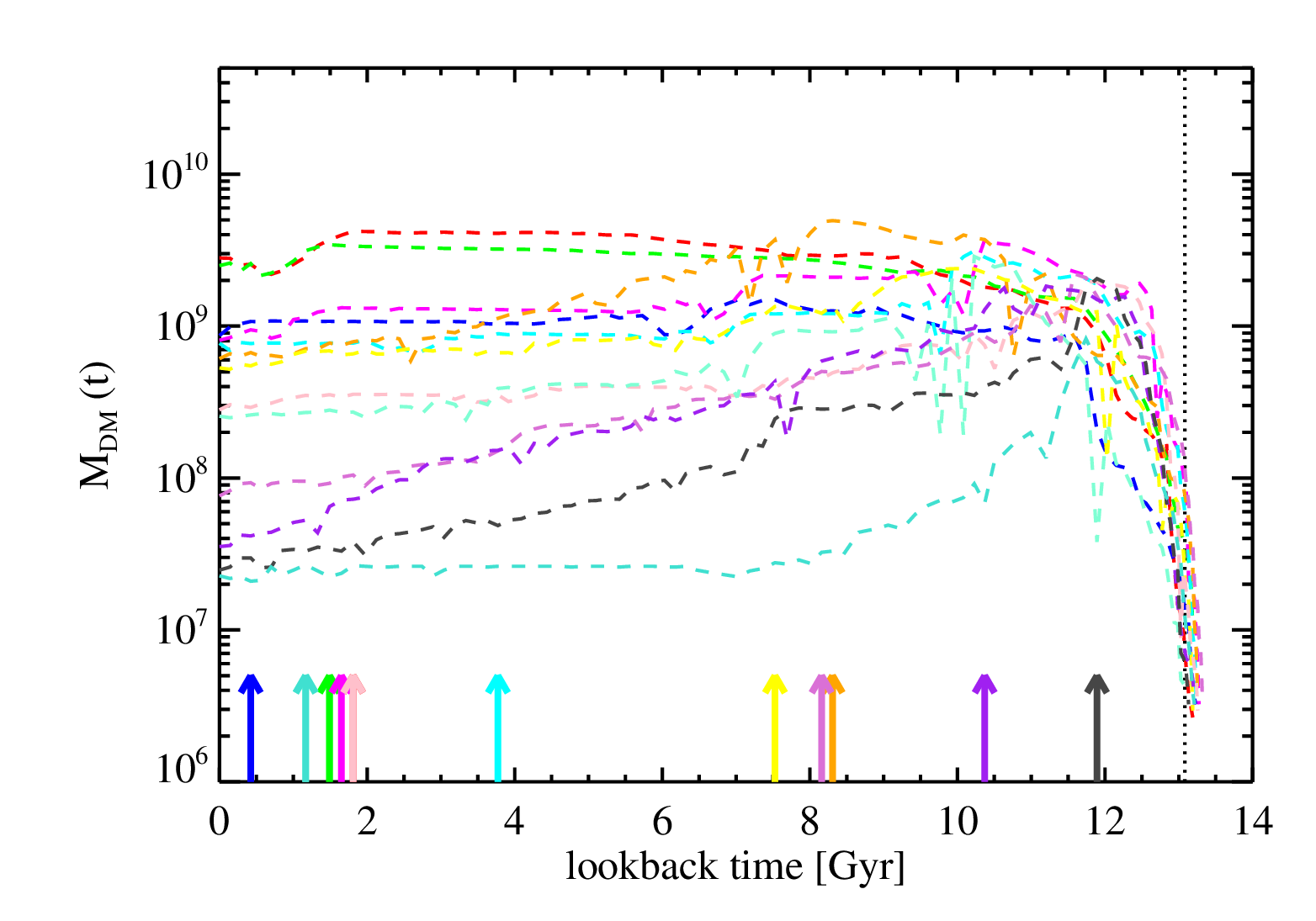}}

{\large $-10<M_{V}<-8$}
\centerline{\includegraphics[width=0.45\textwidth]{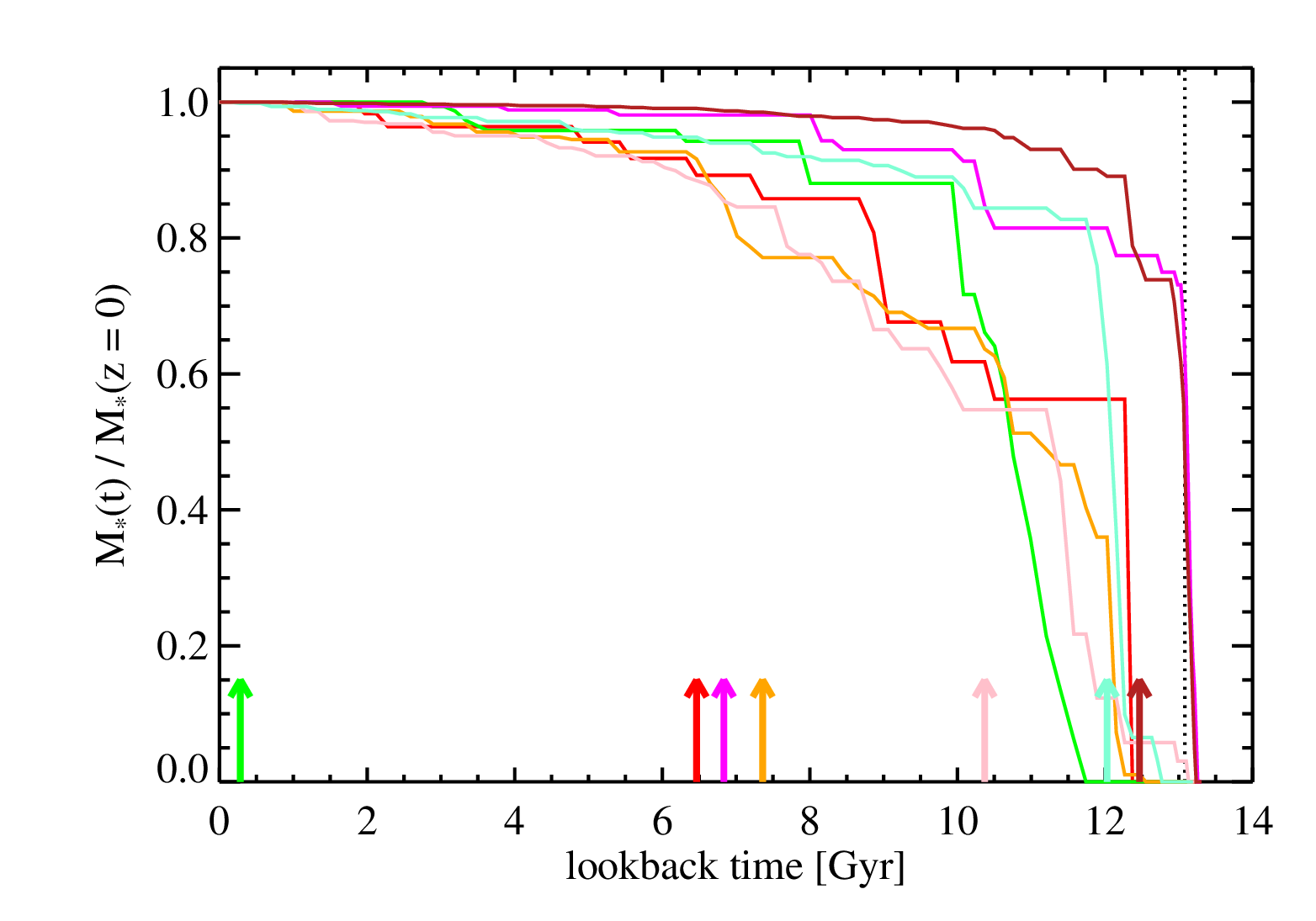}\includegraphics[width=0.45\textwidth]{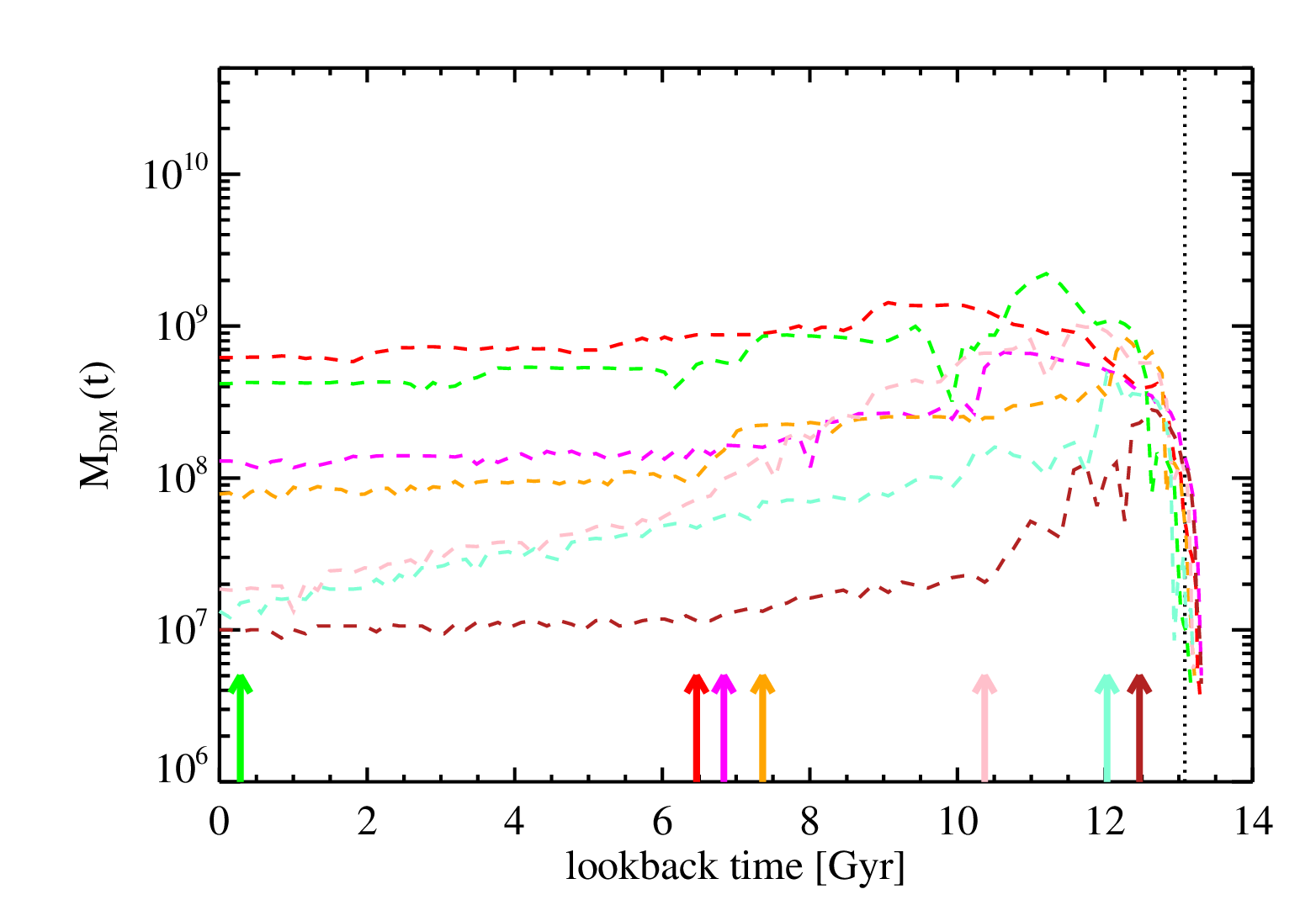}}
   \caption{Evolution of the stellar mass (left) and dark matter mass (right)
     for satellites in the fiducial satellite-model.  Satellites are sorted by
     $M_{V}(z=0)$ into three groups: $-16<M_{V}<-13$ (top panel);
     $-12<M_{V}<-10$ (middle panel); $-10<M_{V}<-8$ (bottom panel). Stellar
     masses are normalised by the present-day values, $\mstar(z=0)$.  Different
     colours correspond to different satellites, and the arrows indicate the
     accretion time, defined as when a satellite was identified as a central
     galaxy for the last time.  The same colour is used to plot the stellar and
     the dark matter mass for a given satellite in the panel aside.  The
     vertical dashed lines mark the end of the reionization, $z_{r}=11.5$, in
     the satellite-model.}
\label{sfh_fig}
\end{figure*}

In \fig\ref{sfh_fig} we present the evolution of the stellar mass and the
{\small SUBFIND} dark matter mass predicted in our fiducial satellite-model.
We remind that this model does not give any satellite fainter than $M_{V}=-4$,
and gives only 2 satellites fainter than $M_{V}=-5$.  We therefore restrict our
comparisons to the classical MW satellites, and we sort them by their
present-day luminosity into three bins: luminous ($-16<M_{V}<-13$, similar to
Sagittarius and Fornax), intermediate ($-12<M_{V}<-10$ like Leo I and
Sculptor), and faint ($-10<M_{V}<-8$ \ie Leo II, Sextans, Carina, UMi and
Draco). Satellites belonging to these three bins are shown from top to bottom
in \fig\ref{sfh_fig}. We do not include here the two model satellites (one in
the luminous and one in the intermediate luminosity bin) which have more
stellar mass than dark matter mass at $z=0$, due to significant tidal
interactions with the main halo. We recall that in our modelling, we do not account
for tidal stripping of stars and for the loss of the cold gas due to
e.g. ram-pressure.  Therefore, we also refrain from considering systems which form $> 50$
per cent of their stars after becoming satellites of the MW-like halo.  This
choice is motivated by the prevalence of old stars (\ie $>$ 10 Gyr) seen in the
MW dSphs \citep{dolphin05,orban08}.  When excluding these galaxies, the number
of satellites is reduced from 16 to 13 in the intermediate bin; 14 to 7 in the
low luminosity bin.  We will later see that faint satellites are mostly
accreted before $z=1$ thus their gas should have been reduced due to the
interactions with the central galaxy and the star formation should have (on
average) ceased shortly after being accreted.

It is encouraging that the model satellites show in \fig\ref{sfh_fig} 
all contain stars older than 10 Gyr regardless of their
luminosities, in good agreement with observations\footnote{In the full sample of model satellites, 43 out of 51 made their first generation of stars $\gta$ 10 Gyr ago.  Those that are not dominated by old stellar populations are excluded in the analysis here with the criterion that half of the stars were in place before the accretion.}.  The most luminous satellites
build up their stellar content over a longer period of time compared to the
faintest ones as observed in the Local Group satellites \citep{dolphin05}.
The five systems that have been accreted earliest ($>$ 9 Gyr), are associated
with fainter satellites ($M_{V}>-12$), and are dominated by old stars, which
means that these galaxies stopped forming stars soon after the accretion.  The
most luminous model satellites are associated with massive dark matter
subhaloes ($\mdm > 10^9\msun$) at present time.  The Figure also shows that all
bright satellites consist of less than 1 per cent of stars made by the end of
the reionization ($z=11.5$ equivalent to a lookback time of $\sim$ 13 Gyr),
while a few of the faint ones contain $> 50$ per cent of such very old stars.
None of the most luminous satellites was accreted before $z\sim1$, and they have
the most extended star formation histories and reside in the most massive
haloes. This is a natural consequence of the fact that subhaloes that were
accreted early, have typically experienced a more significant mass loss than
those accreted later \citep{delucia04a,2004MNRAS.355..819G}. As explained
earlier, no cooling is allowed on satellite galaxies in our model. After
accretion, these galaxies keep forming stars, with the size of their gaseous disk fixed
at the value when they were accreted. The surface density of the gas soon drops
below the density threshold for star formation, therefore rapidly quenching any
further star formation in these galaxies.
  
\subsection{Other properties of the satellites}
\subsubsection{Radial distribution}
\begin{figure}
\centerline{\includegraphics[width=0.5\textwidth]{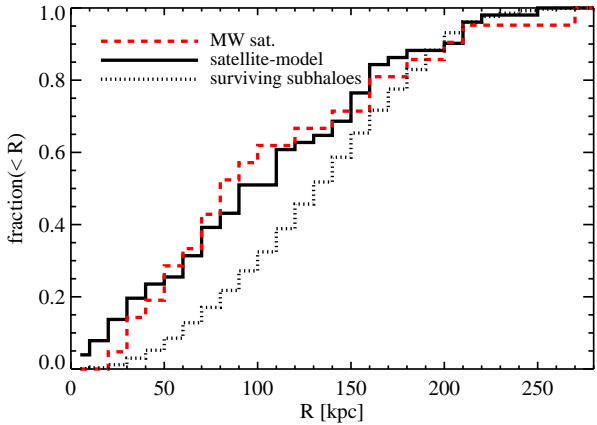}}
   \caption{Normalised cumulative radial distributions of the MW satellites
     (dashed line) compared with that for model satellites (solid line) and for
     the dark matter subhaloes (dotted line) at $z=0$.  The distance from the Sun to the Galactic centre is taken as $\rsun=8.0$~kpc.}
\label{radialdist_fig}
\end{figure}

\begin{figure}
\centerline{\includegraphics[width=0.5\textwidth]{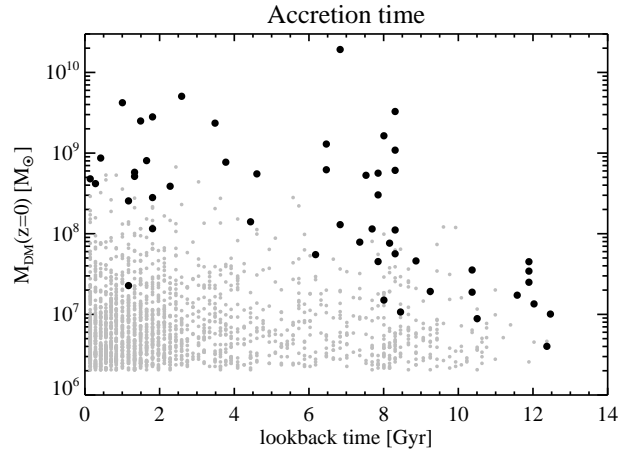}}
   \caption{Present-day bound dark matter mass as a function of the accretion
     time.  Black symbols represent the luminous model satellites, while grey
     symbols mark dark matter subhaloes which failed to form stars.}
\label{dmh_gal_accretiontime_fig}
\end{figure}

\fig\ref{radialdist_fig} shows the cumulative radial distribution of the MW
satellites as a dashed line, and of the model satellites as a solid line.  Here
we only compare our model results with satellites observed within a
galactocentric distance of 280~kpc (\ie excluding Leo T). The median distances
of the two distributions agree well.  For comparison, we also show the
distribution of all surviving dark matter subhaloes (with or without stars) as
a dotted line.  It is clear that the radial distribution of subhaloes is much
less concentrated compared to that of the MW and the model satellites.  The
reason why our model satellites show a cumulative radial distribution similar
to that of the MW satellites can be understood from
\fig\ref{dmh_gal_accretiontime_fig}.  This Figure shows the present-day $\mdm$
as a function of the accretion time.  The small grey circles are for the `dark'
satellites, and the larger black symbols are for those subhaloes which host
luminous satellites.  The bulk of dark satellites (\ie subhaloes without stars)
in the mass range of $10^6\msun < \mdm < 10^7\msun$ have been accreted in the
last two Gyr, and are preferentially found in the outskirts of the MW halo as,
for this mass scale, dynamical friction is not important. Since these small
dark satellites dominate by number, this leads to a much more even distribution
for subhaloes than for the luminous satellites. Half of the satellites were
accreted more than 7 Gyr ago, and the most massive ones (\eg $\mdm >
10^9\msun$) at $z=0$ were all accreted after $z\sim 1$ (see also
\fig\ref{sfh_fig}).  There is also a clear bias for the least massive subhaloes
hosting stars to have been accreted earlier (as mentioned in
\sect\ref{sf_sect}).

\subsubsection{Luminosity-size relation}
\begin{figure*}
\centerline{\includegraphics[width=0.5\textwidth]{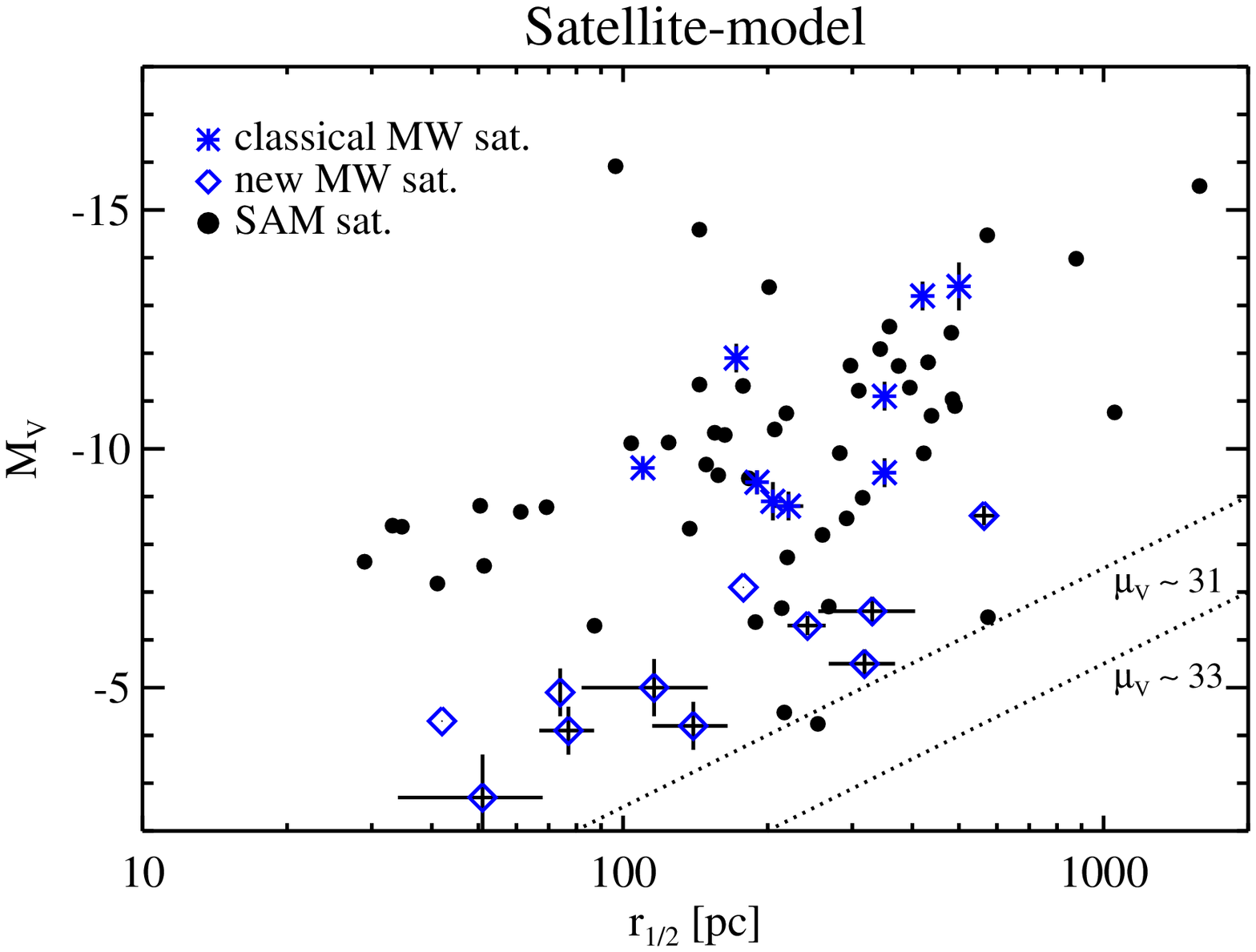}\includegraphics[width=0.5\textwidth]{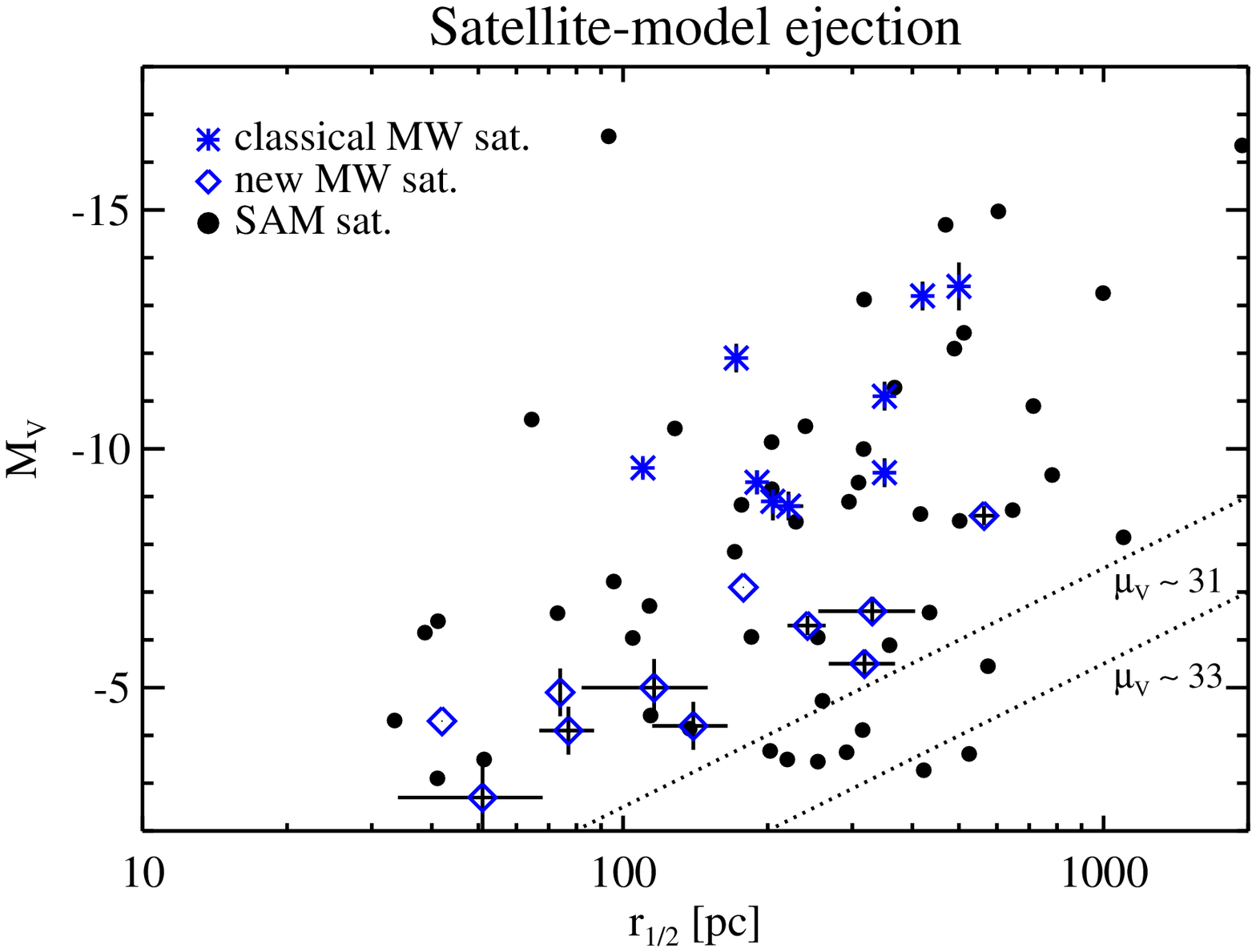}}
   \caption{Luminosity as a function of the half-light radius for the MW and
     model satellites.  Observational measurements are taken from various
     sources: For the classical dSphs, data are from the compilation of
     \protect\cite{vdb00}. Data for the ultra-faint dwarfs are from
     \protect\cite{martin08}, except for Leo V which is taken from
     \protect\cite{belokurov08}.  See the caption of \fig\ref{lumfn_fig} for
     the sources of the total $V$-band luminosity.}
\label{lumsize_fig}
\end{figure*}

Here we compare the half-light radii, $r_{1/2}$, of model satellites with the
observed distribution as a function of the total $V$-band absolute magnitude.
The left panel in \fig\ref{lumsize_fig} is for our fiducial model, while the
right panel is for the alternative feedback scheme. The size of a satellite in
our models is obtained assuming the stars are distributed in an exponential
disk and is roughly proportional to the virial radius of the associated dark
matter subhalo, \ie $r_{1/2} \approx
1.2\lambda\rtwoh$ \footnote{$r_{1/2}\approx 1.68\ r_{D}$ for an exponential
  disk, and $r_{D}$ is the scale length of the disk which is computed as $r_{D}
  \sim \frac{\lambda}{\sqrt{2}}\rtwoh$ to the first order.}, where $\lambda$ is
the spin parameter of the associate DM halo. In \fig\ref{lumsize_fig} we show
the classical dSphs and the ultra-faint satellites with asterisks and diamonds
respectively. The black circles show model satellites. The half-light radii of
the ultra-faint satellites are from the recent study by \cite{martin08} based
on SDSS data, while those of the classical ones are from \cite{vdb00}.

Model satellites with either of the feedback recipes have sizes between $\sim$
30 to 2000~pc, in quite good agreement with the MW satellites, especially the
classical ones.  For the luminous satellites,
  both models extend to larger sizes.  The lack of faint satellites below $M_{V}= -5$ in the fiducial
model makes the comparison with the ultra-faint satellites inconclusive. The compact satellites with $r_{1/2}<100$~pc are a few magnitudes brighter than the
ultra-faint MW satellites of similar size in this model.  

If we compare the sizes of galaxies associated with the same DM subhalo in the two
feedback schemes, those given by the alternative feedback recipe (right panel)
are a bit larger than in the standard feedback model.  This is expected as in
the alternative feedback scheme, more gas is ejected by SNe in small (central)
galaxies and this is reincorporated later, after the halo has grown in mass and
size.  Since the alternative feedback recipe pushes the luminosity function
towards the fainter end, the comparison with the ultra-faint satellites is in
better agreement than in the standard recipe.  In addition, there are some
faint satellites that are larger than the observed ultra-faint systems with
central surface brightness fainter than the current observational limit
$\mu_{v}\sim 30\magsq$ (see dotted lines in the Figure).
\subsubsection{Metallicity-luminosity relation}
\begin{figure}
\centerline{\includegraphics[width=0.5\textwidth]{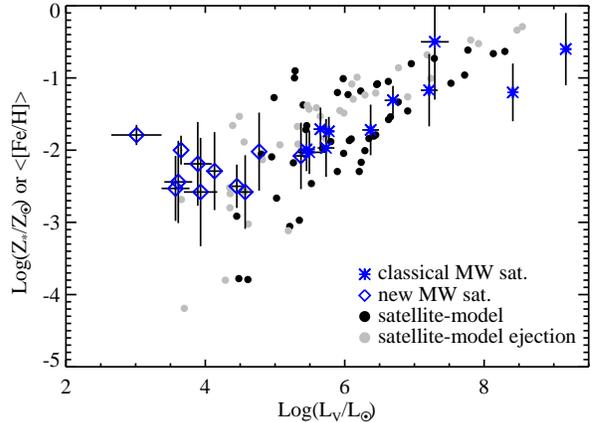}}
   \caption{Metallicity-luminosity relation for model and MW satellites.  Black
     circles show model satellites in the fiducial model, and grey circles are
     used for the \textit{ejection} model.  For the model satellites, the
     logarithmic values of the total metallicity in stars normalised to the
     solar value ($\zsun=0.02$) are plotted.  For the MW satellites, shown in
     (blue) asterisks and diamonds, the mean $\feh$ is used, and error bars 
     denote the dispersions of
     $\feh$ in each galaxy.  See the captions of \fig\ref{lumfn_fig_tab1} and
     \fig\ref{metalfn_fig} for the data sources.}
\label{lummetal_fig}
\end{figure}

The classical MW satellites are known to follow a metallicity-luminosity
relation with metallicity increasing with increasing luminosity \citep{mateo98}.
\fig\ref{lummetal_fig} shows the distribution of model satellites (circles) in
the metallicity-luminosity plane, together with the observational measurements
for classical (asterisks) and new (diamonds) MW satellites.  Black circles are
for the fiducial model with the standard feedback scheme, and grey circles are
used for the \textit{ejection} model with the alternative feedback scheme.
Again, we compare $\zstar/\zsun$ for the model satellites with the averaged
$\feh$ for stars in the MW satellites so care is needed when interpreting these
results. The error bars on the $\feh$ values indicate the spread of the
distributions in each MW satellite rather than measurement uncertainties.  The model satellites in
the fiducial model follow a similar trend as the classical MW satellites, while
it is not possible to compare results from this model to the bulk of
ultra-faint satellites since the model does not predict galaxies with
luminosity less than $L_V \sim 10^{4.5}\lsun$ (see also \fig\ref{lumfn_fig}).
The excess of model satellites seen in the luminosity function
(\fig\ref{lumfn_fig} left panel) around $M_{V}\sim -10$ corresponds to
objects with $\log(\zstar/\zsun)$ in $[-2.0, -1.0]$ dex, and also produces a
`bump' in the metallicity distribution.

With the alternative SN feedback recipe, model satellites also follow the
relation of the classical satellites with some hints for metallicities larger
than observed for $L_{V}>10^6\lsun$.  At the faint luminosity end, there is
better agreement with the ultra-faint satellites, since the alternative
feedback recipe predicts fainter and metal-poor satellites below
$L_{V}=10^5\lsun$.  From our model results, it is not clear whether the
luminosity-metallicity relation has a lower limit since we are affected by the
resolution at the very low mass end.  It is also not clear whether the
predicted relation deviates from the ultra-faint satellites below
$L_{V}=10^5\lsun$.  It is interesting to note that if we follow the ridge of
the observed luminosity-metallicity relation from the high and metal-rich end
down to $L_{V}=10^5\lsun$, the corresponding $\feh$ is $\sim -2.5$ which
happens to be the lower limit for the empirical relation to convert the
equivalent width of \caltr triplet to $\feh$ for RGB stars in dSph galaxies
\citep{battaglia08}.
\subsubsection{Cold gas content}
\begin{figure}
\centerline{\includegraphics[width=0.5\textwidth]{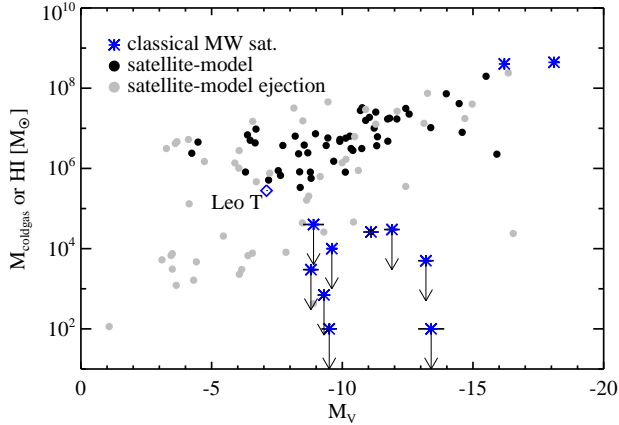}}
   \caption{Cold gas content as a function of $V$-band absolute magnitude.
     Black circles are used for the standard feedback scheme, and grey circles
     for the alternative scheme.  \hone masses are taken from
     \protect\cite{mateo98} for the classical dSphs, and from
     \protect\cite{bruns05} for the Magellanic Clouds.  Apart from the gas rich
     Magellanic Clouds and Sculptor, most of classical dSphs are only
     constrained with upper limits (indicated by the downward pointing arrows).
     The distant Leo T is the only one with a clear detection of $\mhone =
     2.6\times10^5\msun$ among the newly discovered ultra-faint satellites
     \protect\citep{ryanweber08}.  The other ultra-faint satellites have no
     detected \hone or an upper limit of $\mhone \lta 10^3\msun$
     \protect\citep{gp09} and are not plotted here.}
\label{coldgas_fig}
\end{figure}

The majority of the MW dSph satellites (including the ultra-faint ones, see
\citealt{gp09}) are gas deficient.  In \fig\ref{coldgas_fig}, we show the cold
gas content as a function of the luminosity for our model satellites and
compare model results with observational measurements. The Figure shows that
model satellites (from either feedback model) are much more gas-rich than the
observed counterparts, by factors of a few hundred, except at the brightest
end.  The satellites from the alternative feedback recipe populate a similar
region as those from the standard recipe but there are some with significantly
lower (only $\sim 10^{3-4}\msun$) cold gas.  We have verified that altering the
reionization epoch or star formation efficiency does not have a significant
impact on the present-day cold gas mass of satellite galaxies.  This means that
most of the cold gas in satellites was already in place when they were central
galaxies, and that this gas has not been affected much by star formation and/or
feedback. We recall that in our models, star formation only occurs when the mean
surface density of the cold gas is above a certain threshold.  Strong supernova
feedback can in principle reduce the gas content in the cold phase, but most of
the satellites form stars at very low rates, so SN feedback will be less
important. In addition, increasing the supernova efficiency would reduce the
luminosity (and metallicity) of the model galaxies, ruining the agreement with
observational data shown earlier in this study.

It is known that Local Group galaxies show a morphological segregation in the
sense that gas-deficient dSphs are closer to the giant spirals (the MW and
M31), while the gas-rich dIrrs are more evenly distributed \citep{mateo98}.
Recent studies, based on deeper \hone surveys by \cite{gp09} also show that
there is a clear correlation between galactocentric distance and \hone content
for dwarf galaxies in the Local Group.  These are interpreted as indications
that the Local Group environment influences the gas content of satellites
through tidal stripping and ram-pressure stripping \citep[see e.g.][]{ggh03,mayer06}.  
The excessive cold gas associated with our model
satellites might be due to our neglecting of gas removal mechanisms, or to a
simplified calculation of the surface density threshold (particularly for those
satellites which are away from the giant disks, where environment is expected
to play a less important role).
\subsection{Dark matter halo mass and the dynamical properties} 
\label{dmmass_sect}
\begin{figure}
\centerline{\includegraphics[width=0.5\textwidth]{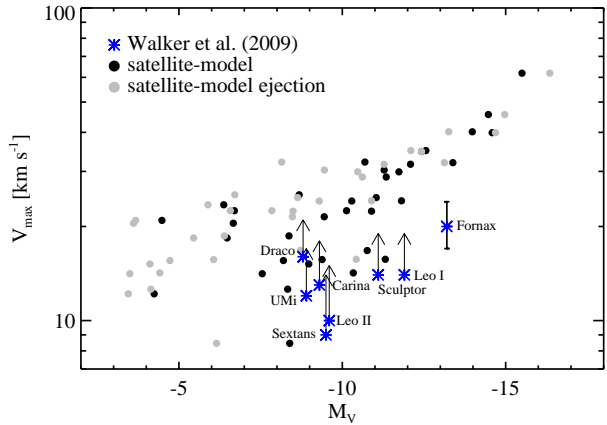}}
   \caption{$\vmax$ as a function of the $V$-band integrated magnitude for the
     non tidally disturbed model satellites.  Black symbols show predictions
     for the fiducial model, and grey symbols are used for the
     \textit{ejection} model.  Data for eight classical MW dSphs are taken from \protect\cite{walker09}.
     The lower limits on $\vmax$ are indicated with the upward arrows except for Fornax dSph whose $\vmax$ is better constrained.}
\label{vmaxlumz0_fig}
\end{figure}

As in \cite{letter}, we trace the evolution of $\vmax$ (\ie the peak circular
velocity of the associated dark matter subhalo), $\mdm$ and the directly measured
$\minnerdm$ in the last 2 Gyrs for each satellite.  16 out of 51 model
satellites are classified as tidally disturbed and these include those two
model satellites which have experienced severe tidal stripping, \ie $\mstar >
\mdm$ at $z=0$ using our fiducial model.  \fig\ref{vmaxlumz0_fig} shows $\vmax$ as a function of $M_{V}$
for those 35 model satellites that do not show signs of tidal disturbance.
Black circles are for the standard feedback scheme, while grey ones are for the
alternative feedback scheme.  The latest constraints on $\vmax$ for 8 classical
MW dSphs based on Jeans/MCMC analysis by \cite{walker09} are shown as
asterisks.  Under the assumption of constant velocity anisotropy, their data could constrain $\vmax$ for Fornax, while only
lower limits were derived for the other dSphs (upwards arrows in the Figure).
Assuming that classical MW satellites are not tidally disturbed, our models
predict that they are associated with dark matter haloes with $\vmax \gta
10\kms$, regardless the choice of SN feedback recipe.  This is fairly
consistent with the constraints by \cite{pmn08} and
\cite{walker09}.

\begin{figure}
\centerline{\includegraphics[width=0.5\textwidth]{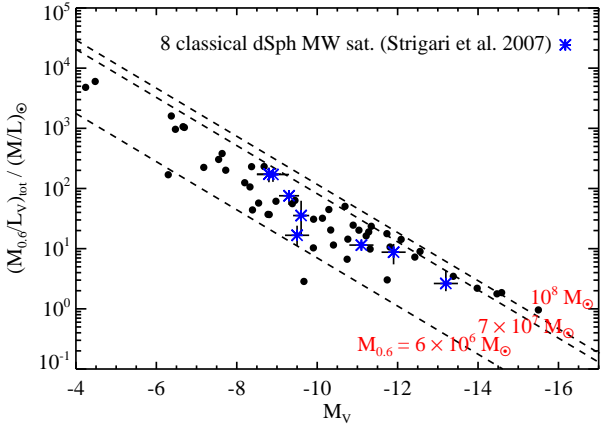}}
   \caption{Mass-to-light ratio of 8 Milky Way classical dSphs and model
     satellites as a function of luminosity.  The mass corresponds to the dark
     matter mass within 0.6~kpc.  For the MW dSphs, data are taken from
     \protect\cite{strigari07} and plotted as asterisks.  The lower and middle
     dashed lines correspond to constant values of $\minnerdm=6\times10^6\msun$
     and $7\times 10^7\msun$, and mark the upper and lower limits of the
     measured $\minnerdm$ for the MW dSphs.}
\label{halomass06_fig}
\end{figure}

\fig\ref{halomass06_fig} shows the mass-to-light ratios calculated using masses
within $0.6$~kpc, $\minnerdm/L$, as a function of absolute magnitude.  The
luminosities shown are from our fiducial model.  The data points for the eight
classical MW dSph (\ie excluding Sgr dSph) measured by \cite{strigari07} are
overplotted as (blue) asterisks.  The lower and middle dashed lines correspond
to constant mass values of $\minnerdm=6\times10^6\msun$ and $\minnerdm=7\times
10^7\msun$ and indicate the upper and lower limits for the observed
$\minnerdm$. For each satellite in our model, two methods can be applied to
measure the $\minnerdm$: 1) to directly sum up bound dark matter particles
within 0.6~kpc from the centre of mass\footnote{determined with the 10 per cent
  most bound particles in each associated subhalo}; 2) to assume that the inner
density profiles are fit by Einasto profiles.  Following \cite{letter}, we
compute $\minnerdm$ by employing the second method for satellites which show
signatures of tidal perturbation\footnote{The two model satellites with $\mstar
  > \mdm$ at $z=0$ are excluded from this analysis.}, and use the direct
summation method for the other satellites.

Our model predicts that the faintest satellites are the most dark matter
dominated, in agreement with observational measurements.  It is encouraging
that the data points and the model satellites follow a similar trend, and that
the spread in the mass-to-light ratio is also comparable.  These results can
translate into the $\minnerdm$-$L_{V}$ plot presented and discussed in
\cite{letter} and are consistent with the claim that the MW satellites are
embedded in dark haloes whose mass is $\sim10^7\msun$ \textit{within the 
optical extent} as proposed by \cite{mateo98}, despite the fact that their
luminosities span nearly five orders of magnitude \citep{strigari08}.  The same 
conclusions hold when we use the luminosities predicted by the alternative 
feedback scheme, and the only difference is that there are more faint 
satellites ($M_{V}>-5$) which have $3,000 <\minnerdm/ L_{V} < 20,000$ [M/L]$_{\odot}$.

Datasets which cover large radius as well as internal proper motions
for stars in these systems are crucial for drawing further conclusions
on the total mass content of the MW satellites.  From our modelling of 
the baryonic physics, we expect a minimum
dark matter halo mass of galaxies before accretion onto a MW-like host
(equivalent $\vvir=16.7\kms$), introduced by the atomic
hydrogen cooling limit (see also the discussion in
\sect\ref{dmh_vs_sat_subsec}).
%
%
%
%---------------------------------------------------------------------------%
\section{Discussion and Implications}
\label{discussion_sec}

\subsection{Number of satellites}

Spectra of distant quasars suggest that reionization was completed by $z=6$
\citep{fan02}, yet the exact duration and processes by which the Universe was
reionized are not well understood.  Our choice of $\zreio=15$ is consistent
with the current observational constraints of $\zreio=11.0\pm1.4$ given by the
\textit{WMAP} 5-year data, which are obtained assuming the reionization
occurred instantaneously \citep{hinshaw09}.  In our model, the suppression of
cooling in haloes with $\vvir < 16.7 \kms$ is the most crucial mechanism which
alleviates the discrepancy in the number of surviving dark and luminous
satellites.  The number of luminous satellites is 1774 in a model that does not
take reionization into account and in which haloes with $\tvir$ lower than
$10^4\kelvin$, are able to cool as much gas as a $10^4 \kelvin$ halo with the
same metallicity.  If cooling is forbidden in these haloes, $N_\mathrm{{sat}}$
reduces to 121. In a model that includes only reionization (no suppression of
the cooling in small haloes) $N_\mathrm{{sat}}=286$ if $\zreio=8$, and
$N_\mathrm{{sat}}=73$ when $\zreio=15$. Forbidding cooling in small haloes, the
number of luminous satellites varies from 88 to 51 when changing $\zreio$ from
8 to 15.  The dependence of $N_\mathrm{{sat}}$ on $\zreio$ in our model appears
slightly stronger than in other recent studies, \eg by \cite{kgk04} and
\cite{maccio09}. Both $N_\mathrm{{sat}}(\zreio=15)$ and
$N_\mathrm{{sat}}(\zreio=8)$ from our modelling are, however, consistent with
current observational constraints, especially when the possible anisotropic
distribution of the satellites is taken into account \citep{tollerud08}.  The
dependence on $\zreio$ found in our model appears to be consistent with recent
results by \cite{busha09}.

Recent hydrodynamical simulations of reionization have suggested that the work
by \cite{gnedin00} might have overestimated the value of the `filtering mass'
\citep{hoeft06,okamoto08}.  We acknowledge that this could imply that our model
may underestimate the number of surviving satellites, especially at low or
intermediate luminosities.  \cite{maccio09} show, however, that effect on the luminosity function of lowering the value
of filtering mass can be compensated with an earlier epoch of reionization. We remind the reader that our study could
not address the scatter in the luminosity function expected as a result of
halo-to-halo variation. \citeauthor{maccio09} show that the number of
satellites changes by about a factor of two in each luminosity bin owing to the
variations of merging histories of host galaxies in the mass range of
$0.8-1.2\times10^{12}h^{-1}\msun$.  Their study also suggests that more massive
central galaxies (with $M=2.63\times 10^{12}h^{-1}\msun$) would host a larger
number of satellites.  These studies can be addressed using new high-resolution simulations of MW-like haloes covering a range of total masses and merging histories \citep[\eg the Aquarius project,][]{springel08}.

Our model does not take into account the loss of stars due to tidal
stripping. We have, however, analysed the evolution of the bound dark matter
mass ($\mdm$) and $\minnerdm$ for the surviving model satellites, and found
that only $\sim 10$ of them show significant evolution in the last 2 Gyrs.
Since baryons are much more concentrated than dark matter, we do not expect
this process to affect significantly the luminosity of our satellites.

Another process that we have not included, and which is potentially important
for the satellite luminosities is ram pressure stripping of the cold gas.  This
is likely to be a stronger effect in our fiducial satellite-model than in the
\textit{ejection} model, where SN feedback itself is strong enough and thus reduces
the cold gas content of satellite galaxies.  The impact of
ram-pressue on the most luminous satellites is also expected to be weak as
these are typically associated with the most massive subhaloes, that were
accreted most recently. A more careful modelling that includes how baryons are
affected by the interactions with the MW-like galaxy is clearly needed
\citep{mayer06}.
\subsection{Satellites associated with small dark matter haloes and ultra-faint
  satellites}
\label{type2_and_uf_subsec}

In our study, we have excluded satellite (Type 2) galaxies whose associated
dark matter subhalo was reduced below the resolution limit of the simulation
(\ie $< 2\times 10^6\msun$).  In our fiducial satellite-model, there are 72
Type 2 galaxies, that cover a wide range of present day luminosities (similar
to that covered by satellites associated with a distinct dark matter
substructure but extended to the fainter end). By tracing their position using the most bound particles of
their associate subhaloes at the last time they were identified, we find that the
median distance of Type 2 galaxies from the host is $\sim 20$~kpc at $z=0$. 90 per cent
of them are within 120~kpc with a maximum distance of $\sim$ 170~kpc, and they
follow a much more centrally concentrated distribution compared to that of Type
1 galaxies \citep[see also][]{gao04}. Among those Type 2 galaxies within the
inner 20~kpc, about 40 per cent are fainter than $M_{V}=-5$ at $z=0$ using the
more effective SN feedback scheme.  If we were to find the
counterparts/remnants of Type 2 galaxies in the observations, they would likely
show a disturbed morphology and/or tidal streams and reside very close to the
host galaxy.  A few ultra-faint satellites which show irregular morphologies
might be indicating they are tidally disturbed, \eg Hercules dSph
\citep{coleman07,sand09}.  However, it is not yet clear whether the
irregularities seen in some ultra-faints are indeed owing to tidal interactions
given the scarce of stars currently observed \citep{martin08}.

\begin{figure}
\centerline{\includegraphics[width=0.5\textwidth]{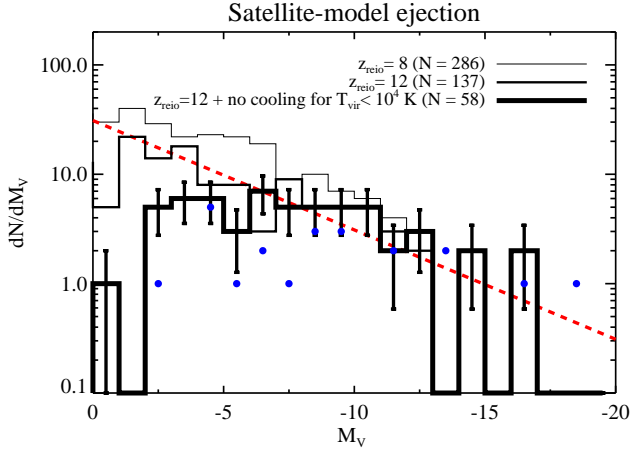}}
   \caption{Luminosity functions for the MW satellites (dashed-line and points)
     and for our model satellites using the more efficient supernova feedback
     scheme (solid histograms).  Data are the same as those in
     \fig\ref{lumfn_fig_tab1}.}
\label{lumfn_molyh}
\end{figure}

The luminosity function predicted by our model is in good agreement with the
power-law-shaped `all-sky' SDSS luminosity function, especially when using the
more efficient SN feedback scheme.  However, this model underestimates the
number of expected ultra-faint (fainter than $M_{V}=-5$) satellites (see
\fig\ref{lumfn_fig} right panel).  Recent theoretical studies have proposed that the
ultra-faint satellites are those fossil satellites which formed before
reionization and managed to cool some gas via $\molehydro$
\citep{br09,sf09,munoz09}.    
In \fig\ref{lumfn_molyh}, we show the luminosity
functions with three different models all using our \textit{ejection} SN
feedback scheme.  The lightest histogram shows results obtained using $\zreio=8$
and allowing haloes with $\tvir<10^4$~K to cool as much gas as a $\tvir=10^4$~K
halo of the same metallicity. The intermediate histogram adopts the same
assumption for cooling in small haloes but assumes $\zreio=12$. The thickest
histogram also adopts $\zreio=12$ but forbids cooling in small haloes.  The two
models using $\zreio=12$ agree well with each other except at the faint end (as
expected). So shutting down cooling in small haloes mostly affects the
luminosity function fainter than $M_{V}=-5$.  If some of those small haloes
could have cooled via molecular hydrogen, they would indeed populate the
ultra-faint satellites regime.  However, it should be noted that, when adopting
$\zreio=12$, there is still a non negligible fraction (about 20 per cent) of
ultra-faint satellites that are associated with haloes with $\tvir>10^4$~K and
whose low luminosities are driven by the very efficient SN feedback.

\subsection{Which dark matter substructures can form stars?}
\label{dmh_vs_sat_subsec}

\begin{figure}
\centerline{\includegraphics[width=0.5\textwidth]{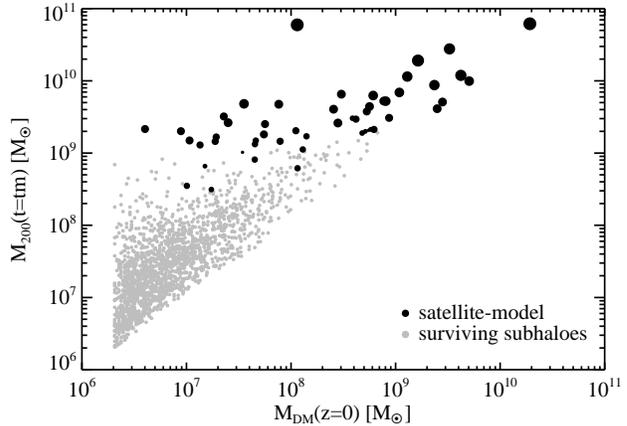}}
   \caption{Maximum $\mtwoh$ mass against present-day bound dark
     matter mass for luminous (black) and dark (grey) satellites.
     The symbol size for the luminous satellites increases with
     the luminosity as given by the satellite-model.}
\label{dmh_gal_masses_tmz0}
\end{figure}

\fig\ref{massfn_res_fig} shows that the mass range $10^8 - 10^9\msun$
is populated with subhaloes both with and without stars in the
fiducial model.  It is interesting to ask why some massive subhaloes
have failed to form stars.  To address this question, we trace the
evolution of the dark matter mass $\mtwoh$ for all satellites and
store the maximum mass and epoch ($t=tm$) when this is reached.
\fig\ref{dmh_gal_masses_tmz0} shows the maximum mass against the
present-day (bound) mass $\mdm$.  Luminous satellites are shown in
black and dark satellites in grey.  The symbol size for the
satellites increases with the luminosity as predicted by the fiducial
satellite-model.  It is clear that most of the subhaloes associated
with the luminous satellites were once much more massive, and that
the most luminous satellites are on average embedded within the most
massive subhaloes at present-day (see also \fig\ref{sfh_fig}).  On
the other hand, dark satellites (in this mass range) have present-day
masses similar to their peak values \citep[see also][]{kgk04}.  It is
also clear that model satellites have a wide range of peak $\mtwoh$
prior to their accretions onto the central galaxy.

\begin{figure}
\centerline{\includegraphics[width=0.5\textwidth]{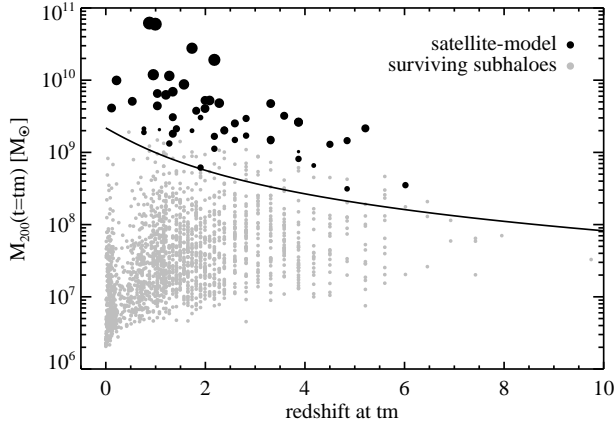}}
   \caption{Maximum $\mtwoh$ mass as a function of the redshift when it was
     reached for luminous (black) and dark (grey) satellites.  The minimum mass
     for cooling via atomic hydrogen at each redshift is indicated by the solid
     line.  Symbols are the same as in \fig\ref{dmh_gal_masses_tmz0}.}
\label{dmh_gal_atmostmassivet}
\end{figure}

We plot the maximum mass as a function of the redshift when this value
was reached in \fig\ref{dmh_gal_atmostmassivet}.  The minimum mass
defined by the cooling via atomic hydrogen as a function of redshift
is indicated by the solid curve.  The Figure shows that most of the
dark satellites were below the threshold and not able to cool gas,
even when they reached their peak mass.  In contrast, luminous
satellites live in subhaloes which have been massive enough and
managed to have sufficient cold gas to fuel star formation.  We also
notice that no luminous satellites achieved their maximum mass before
$z=6$.

We have also checked when and where the present-day luminous satellites
typically formed in the simulation.  We find that the surviving satellites were
all detected in the simulation early, namely their $\mdm$ exceeded $2 \times
10^6\msun$ around the epoch of reionization ($\zreio= 15$ in the
satellite-model) and mostly before the end of reionization ($z=11$), while the
dark ones have emerged throughout the Hubble time.  The distance of the model
satellites to the MW-like galaxy at emergence is typically around $\sim
100$~kpc.

\cite{font06} suggest that surviving satellites were accreted up to 9 Gyr ago
and therefore have been through a different chemical enrichment history
compared to those that were accreted very early on and contributed to the
stellar halo. In our model there are a few objects that became satellites more
than 10 Gyr ago and have survived the tidal interactions with the MW-halo (see
\fig\ref{sfh_fig} and \fig\ref{dmh_gal_accretiontime_fig}).  Typically,
however, their peak masses are smaller than those of the objects that
contributed significantly to the build up of the stellar halo \citep{dlh08}.
It is possible to envision the building blocks of the stellar halo as the most
massive satellites that existed at early times, \ie the counterparts of the very
faint ones, whose orbits did not decay via dynamical friction, and hence have generally been more sheltered from the tidal forces of the Galaxy.
%
%
%---------------------------------------------------------------------------%
\section{Conclusions}
\label{conclusion_sec}
We use a hybrid model of galaxy formation and evolution to study the
satellites of the Milky Way in a cosmological context.  Our method
combines high resolution $N$-body simulations which allow us to trace
the evolution and the dynamics of dark matter haloes directly, and
phenomenological prescriptions to follow the evolution of baryons.
Our adopted semi-analytical recipes and values for the relevant
parameters result in models that reproduce the properties of galaxies
on large scales as well as those of the MW. A few modifications were needed,
however, to reproduce the observational properties of the MW satellites.

With the presence of a reionization background that reduces the baryon content
of subhaloes around $z=15$ and the suppression of cooling for haloes with
$\vvir<16.7 \kms$, our model can reproduce the total number and the luminosity
function observed for the satellites of the MW.  Our fiducial SA model also
shows good agreement with the metallicity distribution and the
metallicity-luminosity relation when a large fraction of newly formed metals in small galaxies are recycled through its hot component and with other properties shared by the MW
satellites, \eg the radial distribution, luminosity-size relation and the star
formation histories.  However, our fiducial model produces an excess of
satellites with $M_{V}\sim -10$ and $\log(\zstar/\zsun) \sim -1$, and does
not predict ultra-faint satellites with the total luminosity below $L_{V} \sim
10^{4.5}\lsun$.

We have tested an alternative SN feedback recipe which is stronger for galaxies
with $\vvir \lta 90\kms$ compared to the standard feedback recipe.  With this
alternative feedback recipe, our model predicts the same number of surviving
satellites (which populate the same set of subhaloes as the standard feedback
model).  The alternative feedback model predicts more satellites with
$M_{V}<-5$ which also follow the metallicity-luminosity relation traced by the
classical and the ultra-faint SDSS MW satellites \citep{kirby08} down to $\feh
\sim -2.5$ and $L_{V} \sim 10^3\lsun$.

Our model satellites are embedded in dark matter haloes with
innermost masses within 600~pc between $6\times 10^6\msun$ and
$7\times 10^7\msun$, in very good agreement with the estimates for the
classical MW dSphs derived by \cite{strigari07}.  This demonstrates
that the existence of a common scale for the innermost mass is a
natural outcome of the CDM galaxy formation and evolution model (see
also \citealt{letter,mkm09,koposov09}). Satellites are dark matter
dominated even within the optical extent, and the mass-to-light ratio
increases with decreasing luminosity. Surviving satellites in our
model are associated with ancient haloes which had masses of a few
$10^6\msun$ by $z\sim 10-20$, and acquired their maximum dark matter
mass after $z\sim 6$.

The brightest satellites in our model are associated with the most
massive subhaloes, were accreted later ($z \lta 1$) and show extended
star formation histories, with only 1 per cent of their stars made by
the end of the reionization (lookback time $\sim$ 13 Gyr). On the
other hand, the fainter satellites tend to be accreted early on,
are all dominated by stars with age $>$ 10 Gyr, and a few of them are
dominated by stars formed before the reionization was complete.  In
our models, the classical MW satellites are associated with dark
matter subhaloes with a peak circular velocity $\gta 10 \kms$, in
agreement with the recent results by \cite{walker09}.

Although our model satellites are in very good agreement with the
latest observations in terms of luminosity, metallicity and the
innermost dark matter content, the agreement with the ultra-faint
galaxies is less conclusive.  Note that the alternative feedback
recipe seems to give a better fit to the properties of ultra-faint
satellites.  This could imply that the ultra-faint satellites of the MW
are associated with dark matter haloes with lower $\vvir$ compared to
the classical ones (\ie they are more sensitive to SN
feedback).  However, until we properly model the loss of baryons due to tidal
stripping and ram pressure in these small systems, and until we obtain
more observational constraints on these objects, their nature will
probably remain unclear.
%
%
%
%---------------------------------------------------------------------------%
\section*{Acknowledgements}
We thank Felix Stoehr for making his simulations available; Eline Tolstoy,
Scott Trager and Simon White for useful comments on an early version of this
manuscript.  The referee, Alexander Knebe, is warmly thanked for a careful reading and suggestions to improve the presentation of the manuscript.  Y-SL is grateful to the Netherlands Foundation for Scientific
Research (NWO), Leids Kerkhoven-Bosscha Fonds (LKBF) and MPA for the financial
support to visit MPA. GDL acknowledges financial support from the European
Research Council under the European Community's Seventh Framework Programme
(FP7/2007-2013)/ERC grant agreement n. 202781. This work was supported by a
VIDI grant to AH from NWO. The computing support and hospitality of MPA are
also acknowledged for the contributions to this work.  This research has made use of NASA's Astrophysics Data System Service.

\bibliographystyle{mn2e}
\bibliography{references} 

\label{lastpage}

\end{document}